\documentclass[lettersize,journal]{IEEEtran}
\usepackage{longtable}
\usepackage{algorithm}
\usepackage{algorithmicx}
\usepackage{array}
\usepackage[caption=false,font=normalsize,labelfont=sf,textfont=sf]{subfig}
\usepackage{textcomp}
\usepackage{stfloats}
\usepackage{url}
\usepackage{verbatim}
\usepackage{graphicx}
\usepackage{titlesec}
\usepackage{multirow}
\usepackage{graphicx}
\usepackage{multirow}
\usepackage{tabularx}
\usepackage{orcidlink}
\usepackage{rotating}
\usepackage{lscape}    
\usepackage{booktabs}
\usepackage{tikz}
\usepackage{comment}
\usepackage{enumitem}
\usepackage{cite}
\usepackage{listings}
\usepackage{rotating} 
\usepackage{amssymb} 
\usepackage{booktabs} 
\usepackage{xcolor} 
\usepackage{tcolorbox}
\usepackage[utf8]{inputenc}
\usepackage{amsmath}
\usepackage[colorinlistoftodos]{todonotes}
\usepackage{threeparttable}
\usepackage{etoolbox}
\usepackage{subfig}

\usepackage{xspace}
\usepackage{hyperref} 
\usepackage[table]{xcolor}
\usepackage{colortbl}

\usepackage{subcaption}   
\usepackage{placeins}     
\usepackage{pdflscape}    

\titlespacing*{\section}
{0pt}{0.5ex plus 0.5ex minus 0.2ex}{0.5ex plus 0.1ex}
\titlespacing*{\subsection}
{0pt}{0.5ex plus 0.5ex minus 0.2ex}{0.5ex plus 0.1ex}
\titlespacing*{\subsubsection}
{0pt}{0.5ex plus 0.5ex minus 0.2ex}{0.5ex plus 0.1ex}

\definecolor{ColorKGE}{HTML}{E0A858}
\definecolor{ColorKCell}{HTML}{482840}
\colorlet{KGETrans}{ColorKGE!30}
\colorlet{KCellTrans}{ColorKCell!30}

\usepackage{booktabs}
\usepackage{multirow}
\usepackage{siunitx}
\usepackage[table]{xcolor}
\usepackage{tabularx} 

\definecolor{speedupblue}{RGB}{0, 102, 204}
\newcommand{\speedup}[1]{\textbf{\textcolor{speedupblue}{#1$\times$}}}

\makeatletter
\renewcommand{\paragraph}{%
  \@startsection{paragraph}{4}%
  {\z@}{1.1ex \@plus 1ex \@minus .2ex}{-1em}%
  {\normalfont\normalsize\bfseries}%
}
\makeatother

\newcommand{\mlkem}{\textsf{ML-KEM}\xspace}
\newcommand{\mldsa}{\textsf{ML-DSA}\xspace}
\newcommand{\slhdsa}{\textsf{SLH-DSA}\xspace}
\newcommand{\hqc}{\textsf{HQC}\xspace}
\newcommand{\falcon}{\textsf{Falcon}\xspace}
\newcommand{\fndsa}{\textsf{FN-DSA}\xspace}

\newcommand{\keccak}{\textsc{Keccak}\xspace}

\newcommand{\horcrux}{\textsc{HORCRUX}\xspace}

\newcommand{\cvxif}{\texttt{CV-X-IF}\xspace}
\newcommand{\xheep}{\texttt{X-HEEP}\xspace}

\definecolor{commentgreen}{rgb}{0,0.5,0}
\definecolor{codeblue}{rgb}{0.1,0.1,0.7}
\definecolor{darkgray}{rgb}{0.4,0.4,0.4}

\begin{document}

\title{HORCRUX: A Complete PQC RISC‑V eXtension Architecture}

\author{
Alessandra Dolmeta\orcidlink{0009-0006-9480-1352}\IEEEauthorrefmark{1}, 
Valeria Piscopo\orcidlink{0009-0006-3552-7162}\IEEEauthorrefmark{1}, Michael Hutter\IEEEauthorrefmark{2}\IEEEauthorrefmark{3}, Maurizio Martina\IEEEauthorrefmark{1}, and Guido Masera\IEEEauthorrefmark{1}\\
\IEEEauthorrefmark{1}Department of Electronics and Telecommunications, Politecnico di Torino, Torino, Italy\\
\IEEEauthorrefmark{2}University of the Bundeswehr Munich, Germany\\
\IEEEauthorrefmark{3}PQShield, Austria
}



\maketitle

\begin{abstract}
This work presents a compact RISC-V extension for Post-Quantum Cryptography (PQC) called \horcrux, which provides a unified Instruction-Set Extension (ISE) supporting all NIST-approved PQC algorithms. \horcrux addresses the difficult trade-off between crypto-agility, high performance, and low resource consumption in constrained environments, a balance typically missing in hardware extensions that focus on limited PQC subsets. By targeting shared kernels across \mlkem, \mldsa, \slhdsa, \hqc, and \falcon, the extension introduces new RISC-V instructions executed by a resource-efficient, tightly coupled coprocessor. This architecture is specifically optimized for embedded systems with strict energy budgets and limited area. Experimental evaluation on a Zynq UltraScale+ FPGA demonstrates speedups of up to 129$\times$ for hash-based, 9$\times$ for lattice-based, and 27$\times$ for code-based schemes, while adding fewer than 21k LUTs and 4.4k FFs. ASIC results from post-synthesis characterization in 65 nm CMOS are also reported, alongside a rigorous power characterization to validate the architecture's energy efficiency. The extension’s modular structure maintains backward compatibility with standard RISC-V cores, offering a scalable solution for deploying PQC on constrained embedded systems.
\end{abstract}


\section{Introduction}
\IEEEPARstart{T}{he} advent of large-scale quantum computers poses a fundamental threat to the security of widely deployed public-key primitives such as RSA and ECC, which are vulnerable to Shor’s algorithm. In response, the National Institute of Standards and Technology (NIST) initiated a multi-year standardization effort to identify quantum-resistant alternatives \cite{nistPQCReport}. The resulting Post-Quantum Cryptographic (PQC) schemes address authenticity and confidentiality through post-quantum Digital Signature (DS) schemes and Key Encapsulation Mechanisms (KEMs), respectively, as efficient alternatives to traditional signature protocols and Public-Key Encryption (PKE). Following years of evaluation, NIST selected a first set of PQC standards in 2024: \mlkem (FIPS 203) 
for public-key encryption and key establishment, \mldsa (FIPS 204)
and \slhdsa (FIPS 205)
for digital signatures (lattice-based and hash-based, respectively).
Although selected for its compact signatures, \falcon is undergoing refinement due to its reliance on high-precision floating-point arithmetic. This dependency complicates constant-time execution and portability, leading NIST to develop an integer-only version, \fndsa (FIPS 206)
before finalization.
More recently, \hqc (Hamming Quasi-Cyclic), a compact and efficient code-based KEM, was also selected for standardization (NIST IR 8545)
reinforcing the evolving PQC landscape.
However, the transition from algorithmic selection to large-scale deployment has shifted the research focus toward the \textit{implementation gap}, where the theoretical security of these primitives must be reconciled with the practical constraints of heterogeneous environments. Despite their security, PQC schemes introduce significant overhead in latency and memory compared to classical primitives, requiring a delicate balance between computational speed, memory footprints, and the substantial communication overhead introduced by larger post-quantum keys and signatures \cite{SAKWA2026100975}. 
While software-only implementations for 64-bit platforms exist \cite{10723802}, they often exceed the capabilities of resource-constrained embedded systems. Consequently, dedicated hardware acceleration is essential to meet stringent performance and area constraints. 
\\\textbf{Related Works.} Instruction-Set Extensions (ISEs) offer a middle ground between pure software libraries and full hardware accelerators, accelerating critical kernels with less overhead. While recent software efforts \cite{cryptoeprint:2024/1515} focus on providing optimized implementations by leveraging standard RISC-V ISA extensions (e.g., Bitmanip or Vector units), our work targets custom hardware acceleration to achieve higher performance across a wider range of primitives. In the landscape of hardware acceleration, loosely-coupled coprocessors have been proposed for specific primitives like lattice-based schemes \cite{Banerjee_Ukyab_Chandrakasan_2019, 8715173, 10776955} or hash-based signatures \cite{9180550, 9460703, 9946370, cryptoeprint:2024/367}. However, these often incur significant communication overhead.
Tightly-coupled ISEs mitigate this by targeting specific PQC bottlenecks. Notable examples include lattice arithmetic for \mlkem and \mldsa \cite{cryptoeprint:2020/049, cryptoeprint:2024/1192,  10294284, 10.1145/3643826, 9605604, cryptoeprint:2023/1505, 10562296, 9609917, 10.1007/978-981-95-8399-7_12}, hash-based acceleration for \slhdsa \cite{9217834, 9460703, Sphincs_hashes, 10830991, 10915711}, or code-based transformations for \hqc \cite{10993202, 11414189, cryptoeprint:2025/601}. Other works have proposed accelerators for specific algorithm pairs \cite{Wang_Zhang_Zhang_Gu_Cao_2024} or broader arithmetic frameworks \cite{10.1145/3579092, 9773945, Ye_Song_Zhang_Chen_Cheung_Huang_2024}.  Notably, support for \falcon in the ISE landscape remains sparse, typically limited to standalone hardware or software optimization \cite{PQC_OpenTitan, 10817843, cryptoeprint:2023/1885}. Although recent designs \cite{s23239408} cover multiple NIST finalists via the \cvxif interface, they often lack end-to-end benchmarks or comprehensive support across all standardized families.
As summarized in \autoref{tab:scheme-comparison}, existing works generally focus on a subset of primitives or specific algorithm families. In contrast, we propose a unified architectural framework that spans the entire NIST PQC portfolio, delivering complete implementation and end-to-end benchmarking for every standardized algorithm family.

\begin{table}[t] 
\centering
\scriptsize
\setlength{\tabcolsep}{3pt}
\renewcommand{\arraystretch}{1.3}
\resizebox{\columnwidth}{!}{
\begin{tabular}{l ccc ccc c cc cc}
\hline
\multirow{2.5}{*}{\textbf{Work}} & \multicolumn{3}{c}{\textbf{ML-KEM}} & \multicolumn{3}{c}{\textbf{ML-DSA}} & \textbf{SLH-DSA} & \multicolumn{2}{c}{\textbf{HQC}} & \multicolumn{2}{c}{\textbf{Falcon}} \\ \cline{2-12} 
 & Sym & Poly & Smpl & Sym & Poly & Smpl & Sym & Sym & Poly & Sym & Poly \\ \hline 

\cite{cryptoeprint:2020/049, cryptoeprint:2023/1505, 9605604} & & \checkmark & & & \checkmark & & & & & & \\ 
\cite{9609917} & & & & \checkmark & \checkmark & \checkmark & & & & & \\ 
\cite{10294284} & & \checkmark & & & & & & & & & \\ 
\cite{10.1145/3643826} & & & & \checkmark & \checkmark & & & & & & \\ 
\cite{9217834, 9460703, Sphincs_hashes, 10915711} & & & & & & & \checkmark & & & & \\ 
\cite{cryptoeprint:2025/601, 10830991} & \checkmark & \checkmark & & & & & & \checkmark & \checkmark & & \\ 
\cite{9946370, cryptoeprint:2024/1192, 10.1007/978-981-95-8399-7_12} & \checkmark & \checkmark & \checkmark & \checkmark & \checkmark & \checkmark & & & & & \\ 
\cite{s23239408} & \checkmark & \checkmark & \checkmark & \checkmark & \checkmark & \checkmark & \checkmark & \checkmark & \checkmark & \checkmark & \checkmark \\ 
\cite{10993202, 11414189} & & & & & & & & \checkmark & \checkmark & & \\ 
\cite{Fritzmann_Sigl_Sepúlveda_2020, 10562296} & \checkmark & \checkmark & \checkmark & & & & & & & & \\ 
\cite{10.1145/3579092} & & & & \checkmark & \checkmark & & & & & \checkmark & \checkmark \\ 
\cite{PQC_OpenTitan, Ye_Song_Zhang_Chen_Cheung_Huang_2024} & \checkmark & \checkmark & \checkmark & \checkmark & \checkmark & \checkmark & & & & \checkmark & \\ 
\cite{cryptoeprint:2023/1885} &  & &  & &  &  & & & & \checkmark & \checkmark \\ \hline 

\rowcolor{gray!10} \textbf{This work} & \checkmark & \checkmark & \checkmark & \checkmark & \checkmark & \checkmark & \checkmark & \checkmark & \checkmark & \checkmark & \checkmark \\ \hline
\end{tabular}
}
\caption{Comparison of related PQC hardware/ISE works.}
\label{tab:scheme-comparison}
\end{table}

\textbf{Contributions.} We present \horcrux, a RISC-V ISE for PQC designed to navigate the challenging trade-off between crypto-agility, high performance, and low resource consumption in constrained environments. Using the Core-V-eXtension InterFace 
\footnote{\url{https://github.com/openhwgroup/core-v-xif}}
\horcrux invokes a coprocessor via custom instructions, requiring minimal microarchitectural changes and maintaining compatibility with unmodified toolchains. Our primary contributions are as follows: 

\begin{itemize}

    \item \textbf{Open-source profiling:} We provide cycle-accurate profiling of PQC workloads using NIST's Known Answer Tests (KATs). All reference materials are open-source, ensuring reproducibility and a reliable foundation for hardware benchmarking.
    
    \item \textbf{Unified and Agile Multi-Family ISE:} A cross-paradigm extension accelerating \mlkem, \hqc, \mldsa, \slhdsa, and \falcon. By prioritizing hardware-level resource reuse, our shared-logic architecture achieves the algorithmic agility required for future-proof fallback mechanisms while maintaining an area-optimized footprint essential for battery-powered embedded platforms. This directly aligns with NIST recommendations to maximize resource density in multi-algorithm implementations~\cite{nist_cswp_39}.
    
    \item \textbf{Comprehensive performance and physical evaluation:} We demonstrate up to 129$\times$ latency reductions with a minimal footprint (20,196 LUTs, 4,429 FFs) on a Zynq UltraScale+ FPGA. Furthermore, we provide a rigorous ASIC characterization in 65nm CMOS ($\approx$116~kGE), evaluating the Area-Time Product (ATP) and energy efficiency to validate the design's suitability for resource-constrained deployments.

\end{itemize}

Following NIST’s evaluation framework~\cite{nist-pqc-faq}, we focus our design on a 32-bit core to balance resource constraints and arithmetic efficiency. Consistent with the RISC-V \texttt{Zk} protection profile~\cite{SCA22}, we do not incorporate masking or active fault-injection defenses.

\textbf{Organization.} The rest of the paper is organized as follows. First, \autoref{sec:Profiling} examines the five PQC algorithms, identifying the computational bottlenecks that motivate our hardware optimizations. \autoref{sec:Design} presents the design methodology and \autoref{sec:horcrux_ise} the implementation details of the introduced custom instructions, followed by a comprehensive overview of the system architecture in \autoref{sec:architecture}. \autoref{sec:Results} evaluates the implementation, providing results for performance, power, and area across ASIC and FPGA platforms. The paper concludes with a discussion of future research avenues in \autoref{sec:discussion} and final remarks in \autoref{sec:conclusion}.

\textbf{Source code.} The source code for all our implementations is available open-source.\footnote{GitHub Repository: \url{https://github.com/vlsi-lab/HORCRUX/tree/locket}.}

\section{Profiling and Bottleneck Analysis}
\label{sec:Profiling}

This section analyzes the performance bottlenecks of the PQC algorithms supported by \horcrux. Our analysis is grounded in cycle-accurate profiling performed on a 32-bit RISC-V core. Rather than focusing on theoretical complexity, we highlight the architectural hotspots that justify our hardware-software co-design.
These bottlenecks are directly addressed by the hardware units described in \autoref{sec:Design}, the impact of which is quantitatively validated via power characterization in \autoref{sec:Results}. To maintain consistency, the specific test names used for power measurements are cited in the following paragraphs between round brackets. The profiling results most pertinent to the architectural instruction set design of \horcrux are summarized in \autoref{tab:pqc_bottlenecks_full_cost}, while the complete profiling dataset is available in the project's open repository.

\textbf{Universal Bottleneck: \keccak Infrastructure.} Across all considered schemes, the \keccak-f[1600] permutation serves as the primary engine for hashing and eXtendable-Output Functions (XOF)\cite{keccakVHDL}. While literature often focuses on the core permutation, our profiling reveals a bottleneck shared between raw transformation and data orchestration. This overhead varies by invocation pattern: 
(\textit{i}) one-shot hashing with fixed padding (\texttt{sha3\_256sb}); 
(\textit{ii}) multi-block absorption where the CPU manually XORs data into the state (\texttt{sha3\_256mb}); 
(\textit{iii}) variable-length XOF operations with interleaved rate iterations (\texttt{shake256}); and 
(\textit{iv}) long-stream generation that typically stalls the pipeline due to repeated state-to-memory transfers (\texttt{multi\_squeeze}). 
The performance penalty of these patterns is significantly compounded by the massive permutation volume required by PQC standards. As evidenced by the invocation counts in \autoref{tab:pqc_bottlenecks_full_cost}, \keccak-f[1600] remains a dominant bottleneck across all families. We address this by offloading the 1600-bit state to a dedicated hardware coprocessor with internal absorption logic. This architectural choice eliminates the manual state manipulation and memory-bus contention that otherwise dominate 32-bit software implementations, effectively reducing the manipulation overhead.

\textbf{Hash-based Signatures (\slhdsa).} The performance of \slhdsa is primarily dictated by the massive volume of hash evaluations performed across thousands of Winternitz One-Time Signature (WOTS+) chains. We focus exclusively on \keccak-based variants because SHA-2 lacks the cross-algorithm hardware reusability found in the SHA-3/SHAKE infrastructure. Our profiling identifies several critical kernels: the underlying hashed message transformations (\texttt{thash}) and Pseudo-Random Function address generation (\texttt{prf-addr}) handle base hashing and PRF outputs; the chain length conversion (\texttt{chain-len}) transforms digests into chain step counts; and the WOTS+ chain generation (\texttt{gen\_chain}) and Merkle tree root calculation (\texttt{compute\_root}) manage the core signature structure. The primary challenge remains the \textit{memory-hashing loop}, where the core must frequently update 32-byte hash addresses and fetch seeds. By utilizing a tightly-coupled \keccak unit that allows for in-place squeezing and direct XORing of the address into the hardware state, we bypass software-level buffer management.

\textbf{Lattice-based Schemes (\mlkem, \mldsa, and \falcon).} The computational overhead of lattice-based algorithms is primarily driven by modular reduction kernels within polynomial arithmetic. While the high-level routines are defined by forward and inverse transforms across different schemes (\texttt{mlkem-poly-ntt/intt}, \texttt{mldsa-poly-ntt/intt}, and \texttt{falcon-ntt/intt}), our profiling demonstrates that the primary cycle sinks are the underlying modular operations. 
In \mldsa-44, for instance, Montgomery reductions and \textit{reduce32} operations combined account for nearly 44\% of the execution cycles, totaling over 100,000 calls. Similarly, \mlkem-512 requires approximately 29,000 Montgomery multiplications, while in \falcon, the arithmetic burden is dominated by modular operations (\texttt{f-mq-montymul}), which exceed 970,000 calls. Standard 32-bit ALUs struggle with the high register pressure during coefficient rearrangement and the multiple cycles required for each reduction. 
\horcrux addresses these via a unified butterfly structure that performs single-cycle arithmetic, reducing instruction counts and register-file traffic.
Beyond core arithmetic, performance is penalized by fine-grained bit-manipulation during polynomial sampling. Profiling identifies the generation of the public matrix (\texttt{gen-matrix}) and Centered Binomial Distribution sampling (\texttt{cbd\_eta2}, \texttt{cbd\_eta3}) as primary hotspots due to intensive shift-and-mask operations. Similarly, \mldsa efficiency is constrained by rejection sampling for secret vectors (\texttt{mldsa-rej-eta}) and bit-shuffling during coefficient unpacking. These routines are inefficient on standard cores due to data-dependent branches and manual byte-stitching.
\horcrux bypasses this via a consolidated sampling unit that offloads lane-based bit-slicing and automated bound-checking.

\falcon presents a unique overhead due to its reliance on floating-point routines and normalization (\texttt{falcon-fpr-norm}), which together account for over 30\% of its cycles.
In alignment with the standardization trend toward integer-only arithmetic, \horcrux intentionally prioritizes integer-compatible \falcon hotspots shared across the lattice family. By targeting core arithmetic, specifically Montgomery multiplications (\texttt{f-mq-montymul}), forward and inverse transforms (\texttt{falcon-ntt/intt}), and normalization (\texttt{falcon-fpr-norm}), we optimize the most future-proof and highest-impact subset of the algorithm. This strategic focus maximizes hardware reusability and avoids over-optimizing floating-point paths that may soon be deprecated. While end-to-end speedup is bounded by the remaining software primitives, this approach ensures a portable, area-efficient solution that balances immediate performance with long-term standards agility.

\textbf{Code-based KEM (\hqc).} \hqc shifts the primary computational bottleneck to large-scale binary polynomial convolution. The vector multiplication (\textit{vec\_mul}) is the overwhelming dominant cycle sink, accounting for the 95\% of the total protocol execution. This operation relies on a sequence of Karatsuba operations (\texttt{karats}) and underlying carry-less multiplications (\texttt{gf-carryless}), which are extremely inefficient on word-oriented ALUs lacking dedicated hardware logic. \horcrux addresses this with a dual-mode multiplier capable of single-cycle carry-less execution. Furthermore, the modular reduction of polynomials (\texttt{gf-reduce}) by the irreducible polynomial $x^8+x^4+x^3+x^2+1$ replaces loop-heavy software with fixed-tap XOR logic. This resource sharing, alongside the use of shared \keccak infrastructure and \texttt{barrett} units, exemplifies the cross-paradigm synergy recommended by NIST to achieve optimal efficiency on constrained, battery-powered platforms \cite{nist_cswp_39}. Finally, over 16,000 \texttt{compare32} calls highlight the orchestration overhead in the decoding stage, where syndrome computation and bit-flipping demand high bit-manipulation throughput.

\begin{table}[h]
\centering
\scriptsize
\setlength{\tabcolsep}{3pt}
\renewcommand{\arraystretch}{1.2}

\resizebox{\columnwidth}{!}{%
\begin{tabular}{l l r r}
\toprule
\textbf{Algorithm (TPC)} & \textbf{Operation} & \textbf{Cycles (\%)} & \textbf{Calls} \\ \midrule 

\multirow{6}{*}{\textbf{\mlkem-512} (\textbf{2.9M})} 
& \keccak-f[1600] & 1,785,637 (61.6\%) & 79 \\
& montg & 406,784 (14.0\%) & 29,056 \\
& barrett & 174,720 (6.0\%) & 13,440 \\
& ntt & 166,552 (5.7\%) & 8 \\
& intt & 119,172 (4.1\%) & 4 \\
& basemul & 50,688 (1.7\%) & 2,304 \\ \hline

\multirow{4}{*}{\textbf{\slhdsa-128-fr} (\textbf{24.7B})}
& \keccak-f[1600] & 5,081,922,902 (20.57\%) & 224,834 \\
& gen\_chain & 573,656,160 (2.32\%) & 770 \\
& wots+ & 316,316,352 (1.28\%) & 22 \\
& compute\_root & 10,077,485 (0.04\%) & 55 \\ \hline

\multirow{6}{*}{\textbf{HQC-1} (\textbf{279M})}
& vec\_mul & 265,516,380 (95\%) & 6 \\
& \keccak-f[1600] & 3,254,832 (1.17\%) & 144 \\
& gf-carryless & 1,519,392 (0.54\%) & 4,998 \\
& compare32 & 233,100 (0.08\%) & 16,650 \\
& gf-reduce & 150,903 (0.05\%) & 5,589 \\
& barrett & 26,865 (0.01\%) & 199 \\ \hline

\multirow{4}{*}{\textbf{\mldsa-44} (\textbf{13.7M})}
& \keccak-f[1600] & 7,707,623 (56.3\%) & 341 \\
& reduce32 & 3,194,880 (23.3\%) & 8,192 \\
& montg & 2,802,820 (20.5\%) & 98,394 \\
& ntt & 438,228 (3.2\%) & 9 \\ \hline

\multirow{4}{*}{\textbf{\falcon-512} (\textbf{230M})}
& FPR\_NORM & 41,907,348 (18.2\%) & 388,031 \\
& FPR & 31,438,117 (13.7\%) & 731,119 \\
& \keccak-f[1600] & 11,369,309 (4.9\%) & 503 \\
& modp\_montymul & 3,426,675 (1.5\%) & 975,905 \\ \bottomrule

\multicolumn{4}{p{\columnwidth}}{\textit{\textbf{Note}}: Percentages relative to Total Protocol Cost. Operations are not mutually exclusive.} \\
\end{tabular}%
}
\caption{Bottleneck analysis across all PQC algorithms.}
\label{tab:pqc_bottlenecks_full_cost}
\end{table}

\section{Design Methodology}
\label{sec:Design}

In this section, we detail the custom instructions and hardware units introduced in our ISE, examining their functional role within the target PQC algorithms and their integration into the processor pipeline. Our design is informed by the comprehensive profiling of PQC schemes discussed in \autoref{sec:Profiling}. This analysis identified recurring architectural hotspots where standard 32-bit instructions exhibit poor efficiency.

\noindent Our methodology addresses the challenging trade-off between crypto-agility, low resource consumption, and high performance, a triplet particularly difficult to satisfy in constrained environments. While many of the works presented in \autoref{tab:scheme-comparison} focus on specialized, standalone accelerators, our objective is to design a crypto-agile PQC accelerator for embedded systems characterized by limited hardware resources and strict energy budgets, such as battery-powered devices. By defining a versatile set of instructions that serve as a general-purpose foundation for PQC, we enable systems to maintain flexibility through fallback mechanisms while ensuring future-proof cryptographic functionality.

To achieve this, we implement a tightly-coupled coprocessor that extends the RISC-V ISA. Unlike a simple functional unit, this coprocessor includes a dedicated internal register file. This allows it to preserve internal state across multiple instruction calls, such as the 1600-bit \keccak state, drastically reducing the cycles spent on redundant memory I/O and lowering the overall energy footprint. To ensure this extension remains practical for general-purpose integration, our methodology is guided by two key constraints:

\begin{itemize}[left=0pt, label={}, itemsep=0.5pt, parsep=1pt]
    \item \textbf{\textsf{Constraint 1:}} Instructions must be RISC-V compliant, adhering to standard encoding formats. We utilize the \texttt{R-type} format for dual-operand operations and the \texttt{R4-type} (three source registers, one destination) to maintain compatibility with unmodified toolchains.
    \item \textbf{\textsf{Constraint 2:}} The hardware design must prioritize area efficiency by maximizing resource sharing across different algorithms to minimize the total silicon footprint, making it suitable for area-constrained embedded platforms.
\end{itemize}

Crucially, \texttt{R4-type} encoding is a native part of the RISC-V specification; thus, our integration requires no structural modifications to the core's decoder or pipeline. By following these principles, we have developed an architectural foundation capable of supporting a wide range of PQC algorithms through a unified datapath.

This strategy fulfills the NIST-recommended approach of primitive sharing, or \textit{accelerator reuse}, where different algorithms share the same internal subroutines to overcome hardware capacity limitations \cite{nist_cswp_39}. Rather than utilizing a wide SIMD or vector-based approach, \horcrux relies on hardware reuse for mathematically similar operations. This strategy allows for highly optimized integration of \mlkem, \mldsa, \slhdsa, \hqc, and \falcon. A primary example is our dual-mode multiplier tree, which is shared between lattice- schemes and code-based to perform both modular integer multiplication and carry-less polynomial convolution. While our methodology emphasizes resource sharing to ensure agility, the current specification of \falcon introduces a unique challenge due to its floating-point dependency. To ensure full algorithm coverage and performance on systems with low operating frequencies, we have integrated a dedicated FPR-unit within \horcrux. This unit provides hardware acceleration for specialized operations, ensuring that the heavy precision and normalization logic do not bottleneck the processor while maintaining the flexibility required by modern PQC standards.

\section{\horcrux Instructions Set Extension}
\label{sec:horcrux_ise}

The \horcrux ISE is integrated as a tightly coupled coprocessor, accessible to the main RISC-V pipeline through a dedicated set of custom instructions. Central to our design philosophy is the aggressive sharing of hardware resources across different PQC families to maintain a minimal area footprint. Rather than deploying isolated, algorithm-specific accelerators, \horcrux is composed of highly versatile functional modules, that all operate around a unified internal register file.

\noindent The following paragraphs detail the architecture of these core datapath components, their specific instruction mappings, as summarized in  \autoref{tab:horcrux_isa}, and the state management strategies that enable efficient cross-algorithm crypto-agility.

\subsection{The \horcrux Register File and State Management}

\begin{figure}[h]
    \centering
    \includegraphics[width=\linewidth]{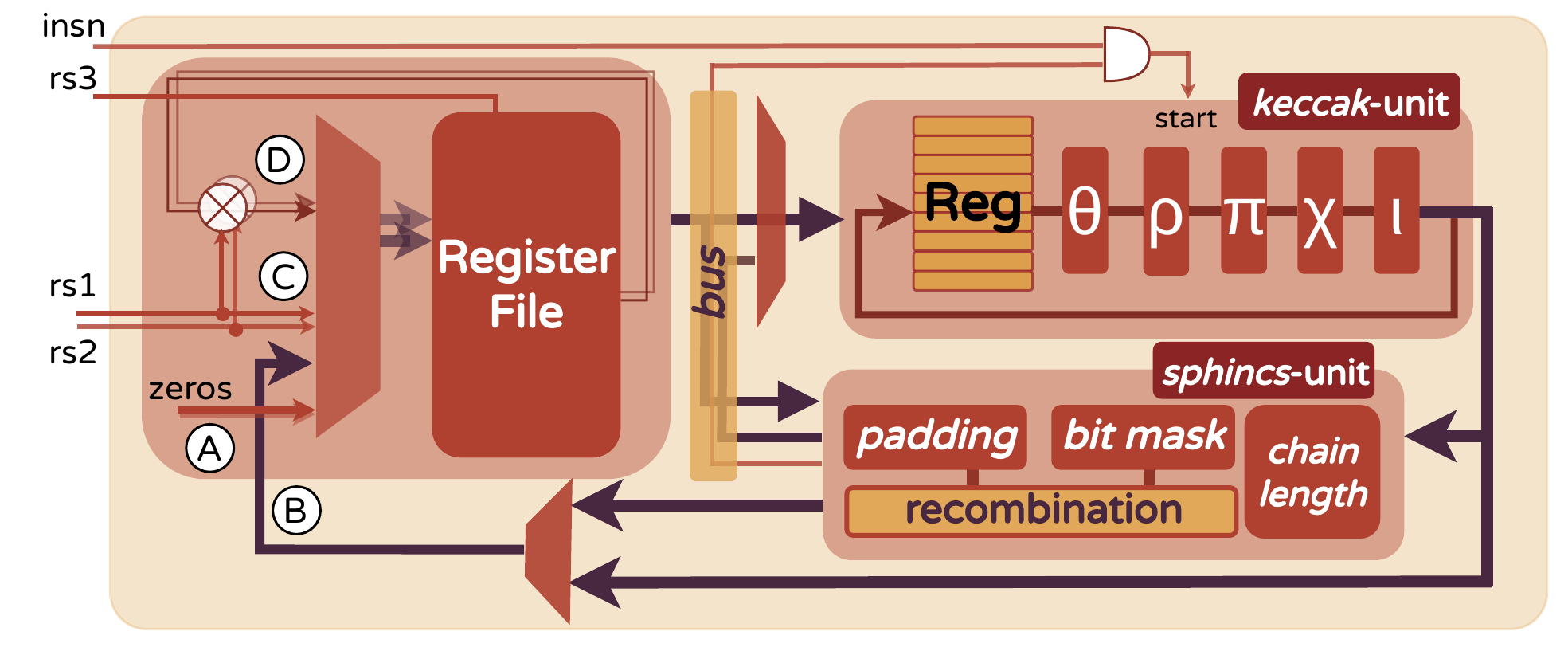}
     \caption{Overview of \horcrux Register File and Hash-Modules.}
     \label{fig:horcrux-hash}
\end{figure}

The foundation of the hash-centric PQC kernels in \horcrux is a unified 50-word by 32-bit register file that exactly mirrors the 1600-bit \keccak sponge state. This design allows the state to persist in hardware across the absorb, permute, and squeeze phases, eliminating redundant memory I/O. The CPU interacts with this internal state primarily through two granular instructions (as detailed in \autoref{tab:horcrux_isa}, \textit{Data Movement \& State Management}): \textbf{\texttt{LOAD}} (an \texttt{R4-type} instruction that writes two 32-bit words from the CPU to the internal RF) and \textbf{\texttt{STORE}} (an \texttt{R-type} instruction that reads a 32-bit word from the internal RF back to the CPU). As illustrated in \autoref{fig:horcrux-hash}, this shared register file supports four distinct write modes beyond asynchronous reset:
\begin{itemize}[left=0pt, label={}, itemsep=0pt, parsep=0pt]
  \item \textbf{(A) \texttt{KINIT}:} A single-cycle instruction that explicitly clears the internal 1600-bit sponge state, replacing 50 individual memory stores.
  \item \textbf{(B) Writeback Mode:} A 50-word bulk update used to commit the state automatically after the \texttt{KPERM}, \texttt{THASH}, or \texttt{PRF} operations complete.
  \item \textbf{(C) \texttt{LOAD}:} Used for filling more than 64 bits of data or during multi-cycle load sequences.
  \item \textbf{(D) \texttt{KABSORB}:} Performs an indexed, in-place XOR injection of new data into the state.
\end{itemize}

This memory architecture drastically reduces instruction overhead. For example, during \slhdsa-128 \texttt{THASH} operations, the hardware expects variables at hardcoded offsets: \texttt{reg[0:3]} for the seed, \texttt{reg[4:11]} for the address, and \texttt{reg[12:15]} or \texttt{reg[12:19]} for the input blocks. While this organization is specific to the 128-bit security level, the coprocessor hardwires alternative register mappings to accommodate the larger state and input requirements of the 192-bit and 256-bit variants. Because the hardware writeback step preserves the address and seed data, the inner loop of the \texttt{gen\_chain} routine only requires the software to reload two words (the \texttt{hash\_addr} field) between consecutive calls, bypassing a full 50-word state reload.
This internal absorption model delivers significant performance gains. Moreover, benchmarks indicate that a SHA3-256 multi-block operation executes in 3,138 cycles, halving the 6,254 cycles required by legacy hardware coprocessors that lack internal absorption. Similarly, evaluating a 67-iteration WOTS+ long chain accelerates from 202,810 cycles to 140,098.

\begin{table*}[!h]
\centering
\caption{The \horcrux Instruction Set Architecture. All instructions utilize standard RISC-V R-type (2 source) or R4-type (3 source) encoding. The naming convention reflects the high degree of reuse.}
\label{tab:horcrux_isa}
\footnotesize
\renewcommand{\arraystretch}{1.0}
\begin{tabularx}{\textwidth}{@{}l l l X@{}}
\toprule
\textbf{Instruction} & \textbf{Type} & \textbf{Algorithm(s)} & \textbf{Description} \\
\midrule \hline
\rowcolor{gray!10} \multicolumn{4}{c}{\textit{Data Movement \& State Management}} \\
\midrule
\texttt{LOAD} & R4 & All & Loads two 32-bit words from CPU to internal RF. \\
\texttt{STORE} & R & All & Reads a 32-bit word from internal RF to CPU. \\
\texttt{COMPARE\_U32} & R & All & Constant-time comparison. \\
\midrule
\rowcolor{gray!10} \multicolumn{4}{c}{\textit{Keccak / SHA-3 Acceleration}} \\
\midrule
\texttt{KINIT} & R & All & Clears the 1600-bit internal register file to zero. \\
\texttt{KSTART} & R4 & All & Triggers 24-round Keccak-$f[1600]$ on internal state. \\
\texttt{KPERM} & R4 & All & Performs in-place permutation with direct writeback. \\
\texttt{KABSORB} & R4 & All & XORs data ($rs1, rs2$) into internal state (absorption). \\
\texttt{KREAD3} & R & All & Reads 3 bytes from state for rejection sampling. \\
\midrule
\rowcolor{gray!10} \multicolumn{4}{c}{\textit{Hash-Based Signatures}} \\
\midrule
\texttt{WOTS\_\{128,192,256\}} & R & SLH-DSA & Computes WOTS+ chains for 128, 192, or 256-bit security. \\
\texttt{PRFADDR\_\{128,192,256\}} & R & SLH-DSA & PRF: Computes $\mathrm{SHAKE256}(sk\_seed \,\|\, addr)$. \\
\texttt{THASH1\_\{128,192,256\}} & R & SLH-DSA & Tweakable hash (robust) for 1-block input. \\
\texttt{THASH2\_\{128,192,256\}} & R & SLH-DSA & Tweakable hash (robust) for 2-block input. \\
\midrule
\rowcolor{gray!10}  \multicolumn{4}{c}{\textit{Unified Multiplier \& NTT Operations}} \\
\midrule
\texttt{BFNTTK} & R4 & ML-KEM & Cooley-Tukey butterfly in $\mathbb{Z}_{3329}$ ($a \pm b\zeta$). \\
\texttt{BFINTTK} & R4 & ML-KEM & Gentleman-Sande butterfly in $\mathbb{Z}_{3329}$. \\
\texttt{BFNTTD} & R4 & ML-DSA & Forward NTT butterfly for Dilithium (low bits). \\
\texttt{BFINTTD} & R4 & ML-DSA & Inverse NTT butterfly for Dilithium (low bits). \\
\texttt{BFNTTDH} & R & ML-DSA & Retrieves high bits of previous Dilithium NTT result. \\
\texttt{BFINTTDH} & R & ML-DSA & Retrieves high bits of previous Dilithium INTT result. \\
\texttt{BFNTTF} & R4 & Falcon & Forward NTT butterfly in $\mathbb{Z}_{12289}$. \\
\texttt{BFINTTF} & R4 & Falcon & Inverse NTT butterfly in $\mathbb{Z}_{12289}$. \\
\texttt{MQMULK} & R & ML-KEM & Montgomery mul: $a \cdot b \cdot R^{-1} \pmod{3329}$. \\
\texttt{MQMULD} & R & ML-DSA & Montgomery mul: $a \cdot b \cdot R^{-1} \pmod{8380417}$. \\
\texttt{MQMULF} & R & Falcon & Montgomery mul: $a \cdot b \cdot R^{-1} \pmod{12289}$. \\
\texttt{MODP\_MONTYMUL} & R4 & Falcon & Montgomery multiplication in mod-$p$ path. \\
\texttt{BARRETT} & R & ML-KEM & Barrett reduction for $\mathbb{Z}_{3329}$. \\
\texttt{RED32} & R & ML-DSA & Conditional reduction for 32-bit signed integers. \\
\midrule
\rowcolor{gray!10} \multicolumn{4}{c}{\textit{Binary Field \& Reductions}} \\
\midrule
\texttt{GFMUL8} & R & HQC & Carry-less $8\times 8$ multiplication over $GF(2)$. \\
\texttt{GF\_REDUCE} & R & HQC & Full $GF(2^8)$ reduction modulo $0x11D$. \\
\texttt{KARATS1} & R & HQC & Karatsuba Step 1: Computes $a_{lo} \times b_{lo}$. \\
\texttt{KARATS2} & R & HQC & Karatsuba Step 2: Computes $a_{hi} \times b_{hi}$. \\
\texttt{KARATS3} & R & HQC & Karatsuba Step 3: Computes cross-term. \\
\texttt{KARATS4} & R & HQC & Karatsuba Step 4: Reconstruction read. \\
\texttt{BARRETT\_HQC\{1,3,5\}} & R & HQC & Barrett reduction for $N \in \{17669, 35851, 57637\}$. \\
\midrule
\rowcolor{gray!10}  \multicolumn{4}{c}{\textit{Sampling \& Rejection}} \\
\midrule
\texttt{CBD\{1,2,3,4\}} & R & \textit{lattice} & CBD sampling for $\eta \in \{1, 2, 3, 4\}$. \\
\texttt{REJ\_UNIFORM} & R & ML-DSA & Rejection sampling mod $Q_{MLDSA}$ for Matrix A. \\
\texttt{REJ\_ETA\{2,4\}} & R & ML-DSA & Nibble-based rejection sampling for secret vectors. \\
\texttt{UNPACK\_Z} & R & ML-DSA & $\gamma_1$-range coefficient unpacking for ExpandMask. \\
\midrule
\rowcolor{gray!10}  \multicolumn{4}{c}{\textit{Falcon FPR Helpers}} \\
\midrule
\texttt{FPR\_LOAD\_SE} & R & Falcon & Loads sign/exponent state for Falcon FPR. \\
\texttt{FPR\_EXEC} & R & Falcon & Executes Falcon FPR packing and returns low 32 bits. \\
\texttt{FPR\_RDHI} & R & Falcon & Reads high 32 bits from previous \texttt{FPR\_EXEC}. \\
\texttt{FPR\_NORM64\_EXEC} & R4 & Falcon & Normalizes 64-bit mantissa and returns low 32 bits. \\
\texttt{FPR\_NORM64\_RDHI} & R & Falcon & Reads high 32 bits from previous \texttt{FPR\_NORM64\_EXEC}. \\
\texttt{FPR\_NORM64\_RDE} & R & Falcon & Reads adjusted exponent from previous \texttt{FPR\_NORM64\_EXEC}. \\
\bottomrule
\end{tabularx}
\end{table*}

\subsection{The \texttt{keccak-unit}}

The permutation itself is offloaded to a dedicated \texttt{\textit{keccak}-unit}, whose core module is 100\% combinational. It computes the five \keccak sub-steps ($\theta$ parity-and-mix, $\rho$ lane rotation, $\pi$ lane permutation, $\chi$ nonlinear step, and $\iota$ round constant XOR) as a 5-step deep logic cloud without any flip-flops, mapping the 1600-bit state in a single clock period. The surrounding datapath holds the registered state and iterates this combinational block 24 times, orchestrated by a 3-state Finite State Machine (FSM).
A fundamental design choice is the separation of the working memory into two distinct buffers: \keccak's \texttt{reg} and the main \texttt{Register File}. While apparently redundant, they resolve inherent timing conflicts. Indeed, \keccak's \texttt{reg} acts as the temporary iterative compute state inside the engine, storing the intermediate matrix between the fully combinational rounds, while the main \texttt{Register File} is the system-visible, architectural shared state that retains data across algorithm phases and instruction boundaries. This dual-register topology decouples the engine's internal pipeline cadence (one update per cycle) from the ISA transaction cadence (instruction-driven partial word writes and selective readbacks). It prevents hazards between internal round progression and external reads, allowing a single \keccak engine to be cleanly time-multiplexed across the different operations.

Building upon this architecture, the execution flow is governed by a highly granular instruction set (detailed in \autoref{tab:horcrux_isa}, \textit{Keccak / SHA-3 Acceleration}). As discussed in the previous section, \textbf{\texttt{KINIT}} and \textbf{\texttt{KABSORB}} manage state initialization and in-place data injection directly within the main \texttt{Register File}. Once the state is prepared, the execution flow cleanly distinguishes between two permutation triggers: \textbf{\texttt{KSTART}} and \textbf{\texttt{KPERM}} (also in \autoref{tab:horcrux_isa}, \textit{Keccak / SHA-3 Acceleration}). \texttt{KSTART} acts as a non-destructive read, processing the state without asserting a writeback, which is ideal when squeezing multiple blocks. Conversely, \texttt{KPERM} is designed for in-place state updates. It asserts a writeback flag one cycle before the permutation begins and clears it upon completion. 

\noindent Finally, to accelerate the rejection sampling required by \mlkem and SPHINCS+, \horcrux introduces the \textbf{\texttt{KREAD3}} instruction (see \autoref{tab:horcrux_isa}, \textit{Keccak / SHA-3 Acceleration}). This single-cycle primitive extracts three consecutive bytes at an arbitrary offset. The hardware forms a 64-bit window from two adjacent register file words, barrel-shifts it by the desired byte offset, and returns the lowest 3 bytes using purely combinational logic. This mechanism eliminates the need for software to read multiple words and manually stitch bytes together across word boundaries.

\subsection{Unified \texttt{sphincs-unit} and Shared-State Orchestration}

At the system level, \slhdsa operations are orchestrated by a dedicated \texttt{sphincs-unit} that leverages the shared \keccak engine. This module is exposed to the CPU via twelve custom instructions (detailed in \autoref{tab:horcrux_isa}, \textit{Hash-Based Signatures}): \textbf{\texttt{WOTS\_\{128,192,256\}}} for chain length calculations; \textbf{\texttt{PRFADDR\_\{128,192,256\}}} for secret key derivation at each security level; and \texttt{THASH1\_\{128,192,256\}} and \texttt{THASH2\_\{128,192,256\}} for robust tweakable hashing on 1- and 2-block inputs at the 128-, 192-, and 256-bit security levels.

We explicitly separate \texttt{THASH1} and \texttt{THASH2}, and provide dedicated variants for each security level, because 1-block and 2-block inputs constitute $\sim$95\% of tweakable hash calls. Each variant uses a fixed register layout whose size scales with the security parameter: 64 and 80 bytes for \slhdsa-128, 96 and 120 bytes for \slhdsa-192, and 128 and 160 bytes for \slhdsa-256. By hardwiring these layouts per instruction, the design avoids the area overhead and control complexity associated with a generic, variable-length instruction or runtime mode bits.
In a standard software implementation, a \texttt{THASH} call would require two distinct SHAKE calls: one to generate a bitmask and a second to compute $\mathrm{SHAKE256}(seed \,\|\, addr \,\|\, (in \oplus bitmask))$. The hardware instead performs the first absorb-and-permute to generate the bitmask, XORs the input, and executes the second permutation internally, eliminating intermediate memory roundtrips. Similarly, each \texttt{PRF} variant absorbs the correctly sized public seed, address, and secret seed in one shot and squeezes the output without any CPU intervention.
Supporting this high-throughput pipeline is the \textit{padding} module (\autoref{fig:horcrux-hash}), which constructs multi-rate padded states entirely in combinational logic. By hardwiring the necessary padding bytes (e.g., \texttt{0x1F} and \texttt{0x80}) for each security-level variant, the architecture eliminates the need for \texttt{memset} or \texttt{memcpy} instructions, streamlining execution of the cryptographic hot loop.
To manage these multi-permutation operations without stalling the processor or requiring multiple \keccak cores, the \texttt{sphincs-unit} employs a streamlined FSM. The execution flows from \textit{IDLE} to \textit{ABSORB\_1}, triggering the first \keccak permutation, and then transitions to \textit{WAIT\_1}. For a \texttt{PRF} operation, the FSM jumps directly to \textit{STORE\_RESULT}. For \texttt{THASH} operations (any security level), the FSM leverages the \textit{bit mask} module (\autoref{fig:horcrux-hash}) to latch the appropriate number of output bytes, 16 or 32 bytes for 128-bit, 24 or 48 bytes for 192-bit, 32 or 64 bytes for 256-bit, from the \keccak output. The \keccak engine is immediately re-armed with the second state for \textit{ABSORB\_2}, effectively time-multiplexing the single \keccak output register to serve as both the intermediate bitmask buffer and the final hash output. Finally, the \textit{recombination} module correctly formats the output for the writeback phase, preserving the address and seed registers to avoid reloading them in subsequent chain steps.

Beyond hashing, the \texttt{sphincs-unit} integrates a \textit{chain length} module to accelerate the pure integer arithmetic required for WOTS+ base-$\omega$ conversion and checksum calculation. Operating on the 32-byte message hash already present in \texttt{reg[0:7]}, the module combinationally extracts 64 little-endian nibbles.
Rather than introducing complex control logic or dedicated sequential datapaths for each security level, \horcrux adopts a highly efficient combinational approach. The checksum calculations, $\sum(15 - nibble[i])$, for the 128, 192, and 256 parameter sets are resolved within a lightweight, shared adder tree network. Because basic integer addition requires a minimal silicon footprint, computing these checksums purely in combinational logic is strictly more area-efficient than instantiating the state machines, sequential control, and intermediate registers that would be required to iteratively multiplex a single adder. No hardware reconfiguration or mode bits are required; the parameter set is entirely dictated by the firmware's choice of \texttt{WOTS} variant.

\noindent The hardware orchestration yields a significant impact: a complete \texttt{THASH1} call requires only ~67 instructions (including memory I/O and the hardware trigger). In contrast, the software equivalent involves two full \texttt{shake256()} calls and extensive state-to-memory transfers, costing hundreds of instructions and causing substantial performance degradation.

\subsection{Unified Multiplier Tree}

All polynomial and field arithmetic across \mlkem, \mldsa, \falcon, and \hqc converge into a single, highly flexible datapath: the \texttt{multiplier\_tree}. Rather than instantiating dedicated arithmetic logic units for each algorithm, \horcrux achieves broad function coverage through instruction-level multiplexing (detailed in \autoref{tab:horcrux_isa}, \textit{Unified Multiplier \& NTT Operations} and \textit{Binary Field \& Reductions}). 

\begin{figure}[h]
    \centering
    \includegraphics[page=1, width=\linewidth]{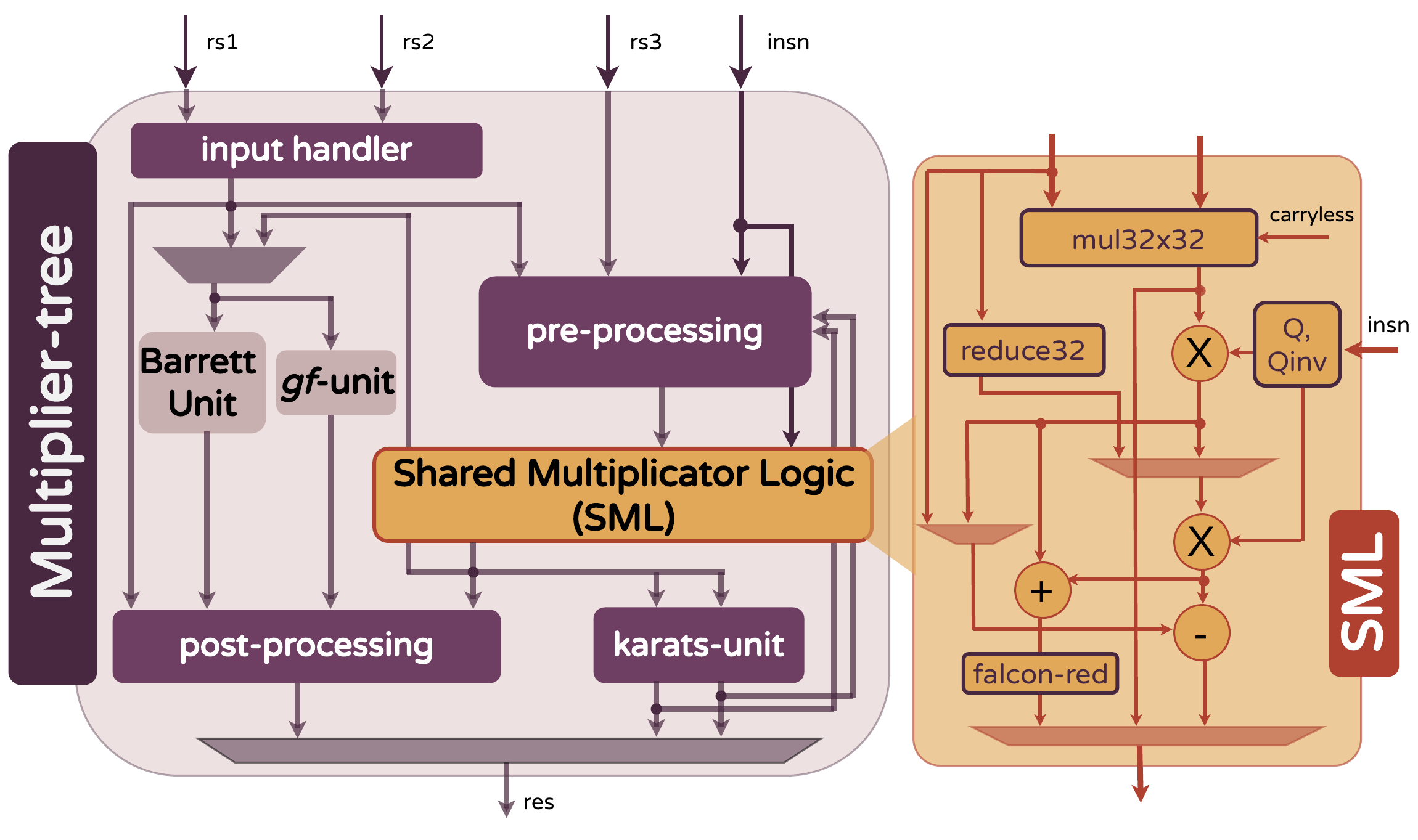}
     \caption{Overview of the Unified Multiplier Tree.}
     \label{fig:multiplier-tree}
\end{figure}

The architectural core of this datapath is the Shared Multiplicator Logic (SML), illustrated in \autoref{fig:multiplier-tree}. The primary novelty of the SML is the \texttt{mul32x32} block, which operates seamlessly across two distinct algebraic domains. By toggling a single hardware \texttt{carryless} control signal, this multiplier switches between standard carry-propagating integer multiplication (for the modular arithmetic in lattice-based schemes) and carry-free XOR accumulation (for the $GF(2)$ polynomial multiplication in \hqc). This dual-mode capability eliminates the need for a separate CLMUL (carry-less multiplication) datapath, securing a massive area reduction. In carry-less mode, the same \texttt{mul32x32} block serves both the \texttt{GFMUL8} instruction---which performs an 8-bit carry-less multiplication of two $GF(2^8)$ elements, producing a 16-bit intermediate later reduced by \texttt{GF\_REDUCE}---and the four-step \texttt{KARATS1}--\texttt{KARATS4} Karatsuba sequence, which decomposes a full 64-bit carry-less multiplication into three 32-bit carry-less multiplications to compute HQC's 128-bit products entirely through the shared hardware.
For lattice-based schemes, the SML implements a universal Montgomery reduction skeleton. Following the initial multiplication of \texttt{mul32x32}, two subsequent hardware multipliers within the SML compute $t = \text{low}(a \cdot b) \cdot Q^{-1}$ and $t \cdot Q$. Instruction-driven multiplexers automatically supply the correct hardcoded parameters ($Q$ and $Q^{-1}$) for \mlkem ($3329$), \mldsa ($8380417$), or \falcon ($12289$), enabling the exact same datapath to execute \texttt{MQMULK}, \texttt{MQMULD}, and \texttt{MQMULF}. Furthermore, the \texttt{MODP\_MONTYMUL} instruction extends this path to accept dynamic prime values at runtime, natively supporting \falcon's complex key generation routines.

\paragraph{Routing and Pre-processing.} Surrounding the SML is a flexible routing network designed to format inputs and outputs for specific cryptographic transforms. The \texttt{input handler} and \texttt{pre-processing} modules (\autoref{fig:multiplier-tree}) compute the necessary sums, subtractions, and XORs required before multiplication, setting up the operands for forward and inverse NTT butterflies (\texttt{BFNTTK}/\texttt{BFINTTK}, \texttt{BFNTTD}/\texttt{BFINTTD}, \texttt{BFNTTF}/\texttt{BFINTTF}).
The \texttt{input handler} integrates internal registers to XOR incoming data with previously latched values, enabling hardware-level accumulation. By managing these stateful transformations locally, the unit conditions operands immediately before multiplication, eliminating the need for intermediate memory cycles or software-managed buffers.
After the SML completes the multiplication, the \texttt{post-processing} module (\autoref{fig:multiplier-tree}) finalizes the butterfly outputs via addition, subtraction, or concatenation. \\Operations that do not require the full SML multipliers are routed through lightweight, dedicated sub-units to further minimize area.

\begin{itemize} 
    \item The \texttt{Barrett Unit}: This module is shared between \mlkem and \hqc. To keep the silicon footprint minimal, the unit avoids using full multipliers. Instead, it implements the reduction logic using a sequence of hardwired shifts and additions. For \mlkem, it performs a constant-time reduction modulo $3329$ (\texttt{BARRETT}), while for \hqc, it supports parameters for 128, 192, and 256-bit security levels ($N=17669, 35851, 57637$, respectively \texttt{BARRETT\_HQC\{1,3,5\}}).
    
    \item The \texttt{reduce32} Module: Integrated within the SML for \mldsa, this module implements a specialized reduction modulo $8380417$. It first computes a correction factor $t = (input + 2^{22}) \gg 23$ and then leverages the SML's auxiliary multiplier to perform the final $t \cdot Q$ subtraction (\texttt{RED32}). 
    
    \item \texttt{\textit{gf}-unit}: Specifically for \hqc's $GF(2^8)$ operations, the unit implements a dedicated reduction instruction. It reduces 16-bit polynomials modulo $x^8 + x^4 + x^3 + x^2 + 1$ ($0x11D$). The logic utilizes a two-pass XOR-based reduction with fixed feedback taps at positions $\{4, 3, 2, 0\}$, ensuring a single-cycle reduction without the overhead of a general-purpose divider (\texttt{GF\_REDUCE}).
    
\end{itemize}

\noindent The SML includes the \texttt{COMPARE\_U32} instruction for constant-time 32-bit equality comparisons. To prevent timing side-channels inherent in branch-based logic, the hardware implements the branchless operation: $rd = 1 \oplus (((rs1 - rs2) \mid (rs2 - rs1)) \gg 31)$. This fixed-latency primitive ensures secure, data-independent verification across all supported PQC schemes (\autoref{tab:horcrux_isa}, \textit{Data Movement \& State Management}).

\noindent Crucially, to maintain strict adherence to the 32-bit register file constraint while computing wide intermediate arithmetic, \horcrux relies on stateful micro-protocols. For \mldsa, the butterfly operations produce outputs wider than 32 bits. Therefore, \texttt{BFNTTD} and \texttt{BFINTTD} return the low half of the result directly to the CPU while latching the complementary high half in an internal register. A subsequent \texttt{BFNTTDH} or \texttt{BFINTTDH} instruction retrieves this stored value. 
Similarly, the 64-bit carry-less multiplication required by \hqc is decomposed into a four-step instruction sequence (\texttt{KARATS1} through \texttt{KARATS4}). By splitting 64-bit operands $A$ and $B$ into four 32-bit halves, the hardware computes the 128-bit product using three 32-bit carry-less multiplications: $P_0 = A_0 \cdot B_0$, $P_2 = A_1 \cdot B_1$, and the cross-term $P_{mid} = (A_0 \oplus A_1) \cdot (B_0 \oplus B_1) \oplus P_0 \oplus P_2$. Because the resulting 128-bit product exceeds the standard CPU register width, the protocol handles the output chunk-by-chunk:
\texttt{KARATS1} and \texttt{KARATS2} compute and return the low ($P_0[31:0]$) and high ($P_2[127:96]$) words while storing intermediates. \texttt{KARATS3} and \texttt{KARATS4} finalize the reconstruction of the middle 64 bits and return them in two 32-bit words.

\noindent This strategy trades a minimal number of CPU fetch cycles for massive hardware reuse, completely avoiding the silicon overhead of 64-bit temporary buses or widened CPU register ports. While this approach maximizes area efficiency, integrating pre-processing, modular reduction, and post-processing into a single-cycle butterfly creates the primary frequency bottleneck of the architecture. This long combinatorial path was a conscious design trade-off to prioritize a shareable, area-efficient datapath over peak operating frequency.

\subsection{fpr-unit}

The specialized arithmetic requirements of \falcon are handled by a dedicated \texttt{fpr-unit}. This module provides hardware acceleration for the complex 64-bit floating-point representation (\texttt{FPR}) and normalization routines that otherwise dominate the algorithm's cycle count. The \texttt{fpr-unit} employs stateful micro-protocols to process 64-bit data through a 32-bit RISC-V interface, utilizing a suite of six instructions (see \autoref{tab:horcrux_isa}, \textit{Falcon Mod-p \& FPR Helpers}).
The formation of a 64-bit floating-point value is split into three phases to accommodate the sign, exponent, and mantissa. The process begins with \texttt{FPR\_LOAD\_SE}, which latches the sign bit and the unbiased exponent into internal registers. Subsequently, \texttt{FPR\_EXEC} accepts the 64-bit mantissa (as two 32-bit words) and triggers the \textit{fpr\_pack64} function. This logic performs critical conditioning, including adding a bias of 1076, clamping the value to zero for exponents below the subnormal threshold, and executing a round-to-nearest (ties-to-even) algorithm based on the lowest three mantissa bits. The instruction returns the lower 32 bits of the result immediately, while the high 32 bits are stored in a register for retrieval by the \texttt{FPR\_RDHI} instruction.
For the high-precision requirements of \falcon's key generation, the unit implements a dedicated normalization datapath. The \texttt{FPR\_NORM64\_EXEC} instruction accepts a 64-bit unnormalized mantissa and an associated exponent. The hardware utilizes a leading-zero count (LZC) module and a barrel shifter to align the mantissa into the $[2^{63}, 2^{64}-1]$ range. The shift amount is simultaneously used to adjust the exponent, ensuring the numerical value remains constant. To maintain the 32-bit register file constraint, the normalized 64-bit mantissa and the adjusted exponent are distributed across three read operations: the lower 32 bits of the mantissa are returned by the execution trigger, while the high 32 bits and the updated exponent are read back via \texttt{FPR\_NORM64\_RDHI} and \texttt{FPR\_NORM64\_RDE}, respectively.
This stateful architecture allows \horcrux to support the full complexity of \falcon's floating-point operations while reusing the core 32-bit bus and register file infrastructure.

\subsection{Sampler}

The Sampling-unit provides a high-speed hardware path for the non-deterministic sampling and unpacking routines essential to \mlkem and \mldsa. It centralizes Centered Binomial Distribution (CBD) sampling, where independent $\eta$-bit populations are summed and subtracted to produce a centered value in $[-\eta, \eta]$, accelerated via four dedicated instructions (\texttt{CBD1}--\texttt{CBD4}). By instantly extracting bits from pre-processed random words, the hardware returns sign-extended coefficients in a single cycle, replacing branch-heavy software logic and minimizing silicon footprint. It further automates rejection sampling for matrix A (\texttt{REJ\_UNIFORM}) and secret vectors (\texttt{REJ\_ETA2/4}), performing modulus checks and nibble-based transformations while returning a result with an integrated validity flag. Finally, the \texttt{UNPACK\_Z} instruction accelerates mask expansion by extracting and range-shifting 18-bit coefficients. This integrated approach creates a memory-free pipeline from the \texttt{keccak-unit} directly to the lattice processors, ensuring that the cryptographic hot loop is limited by hash throughput rather than data orchestration overhead.
\\\newline
\horcrux exposes its capabilities through \textbf{56 custom instructions}. The ISA is designed to be granular, providing primitives that allow software to flexibly compose full protocols. In practice, these instructions are invoked via assembly macros within the original software implementations, substituting performance-critical C functions with direct hardware-accelerated calls. \autoref{tab:horcrux_isa} details the complete instruction set, categorized by functionality.

\section{\horcrux Architecture}
\label{sec:architecture}

\begin{figure*}[h]
\centering
\includegraphics[width=0.9\linewidth]{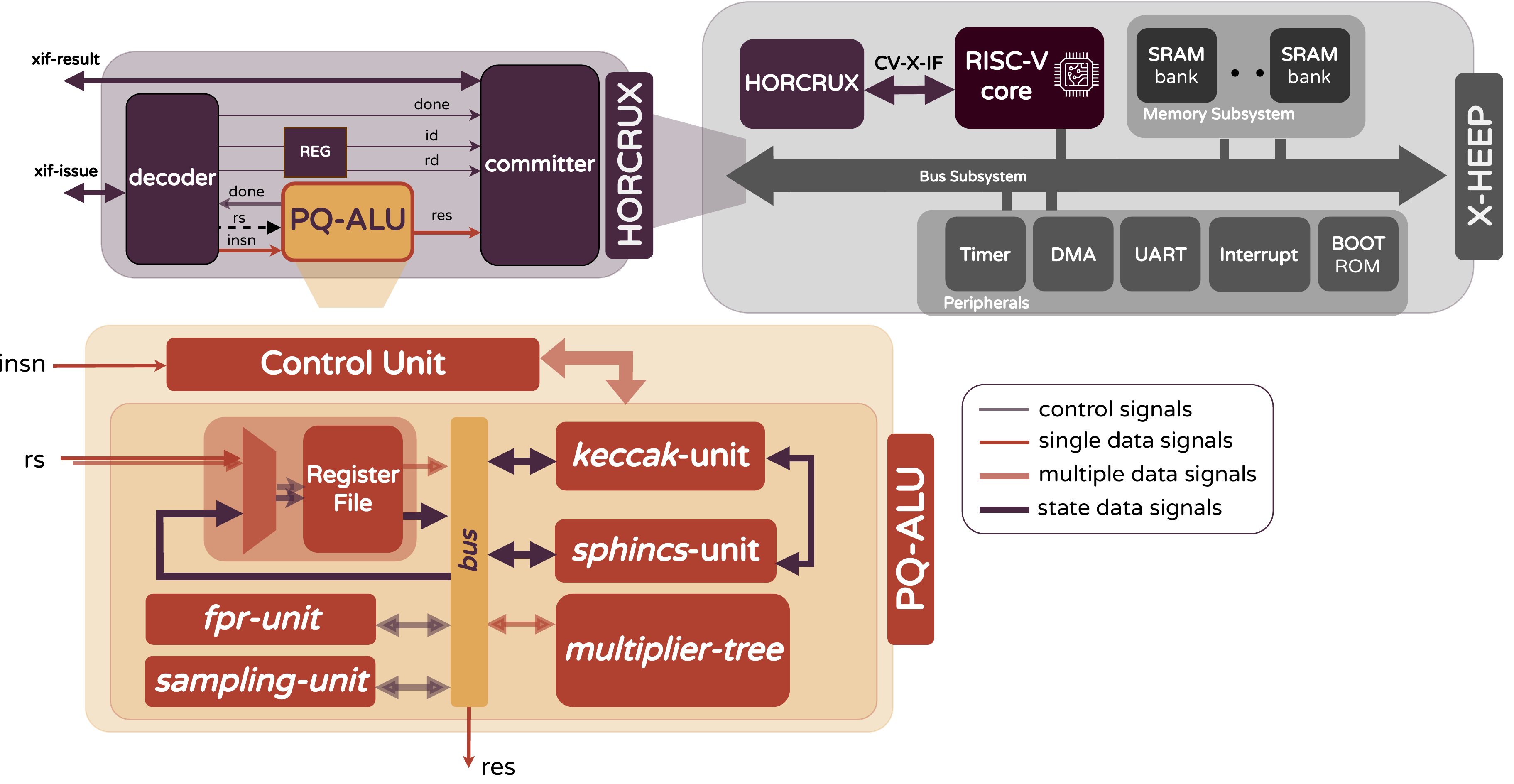}
\caption{Overview of \horcrux architectural layout, signal classification and overview of the integration into \xheep.}
\label{fig:horcrux}
\end{figure*}

The \horcrux coprocessor is a high-performance, tightly coupled accelerator designed for the CV32E40PX RISC-V core. It leverages the Core-V eXtension Interface (CV-X-IF) to achieve seamless integration with the host processor's pipeline. Unlike traditional memory-mapped accelerators, which often require complex DMA transfers and suffer from significant bus contention, \horcrux functions as a native extension of the CPU's execution stage. By intercepting custom instructions directly from the fetch stream, it performs specialized post-quantum (PQ) arithmetic and returns results to the general-purpose registers (GPRs) or updates its internal state with minimal cycle overhead.
The architectural layout is illustrated in \autoref{fig:horcrux}.  For clarity, signals are categorized by their bit-width and function: \texttt{control signals} manage synchronization; \texttt{single data signals} represent standard 32-bit scalar operands; \texttt{multiple data signals} denote multiple 32-bit paths; and \texttt{state data signals} carry high-dimensional internal state.

\noindent \textbf{Coprocessor Integration.} The integration of \horcrux within the host system follows a strictly modular design philosophy, ensuring that the specialized hardware can scale alongside the evolving requirements of PQC without becoming a bottleneck. To achieve this, \horcrux interfaces with the host CPU through a structured three-stage pipeline that mirrors the core's own execution flow while remaining physically and logically distinct. This separation is bridged by the Core-V eXtension Interface (CV-X-IF), a standardized protocol designed by the OpenHW group
that allows for "\textit{plug-and-play}" instruction set extensions. By offloading complex PQC kernels at the instruction level, \horcrux avoids the latency penalties of the system bus and the software overhead of traditional drivers.
\\This modular three-stage flow is organized as follows:

\begin{itemize}[left=0pt, label={}, itemsep=0.5pt, parsep=1pt]
    \item \textbf{Instruction Decoding:} The \texttt{decoder} serves as the primary entry point for the coprocessor. It continuously monitors the \texttt{xif-issue} interface to intercept custom R-type (\texttt{0x3B}) and R4-type (\texttt{0x6B}) opcodes. Acting as a stateless routing controller, this stage parses the incoming 32-bit instruction word, extracts the source register values (\texttt{rs1}, \texttt{rs2}, and \texttt{rs3}), and dispatches them to the appropriate functional unit within the \texttt{PQ-ALU}. Because the instruction mapping is localized within a SystemVerilog package, the ISA can be extended or modified by simply adding a new case entry, requiring zero modifications to the core CPU logic.
    
    \item \textbf{PQ-ALU Execution:} Decoded instructions are executed within the \texttt{PQ-ALU}, which consolidates the arithmetic and logical units described in \autoref{sec:Design}. These modules are within a unified hardware wrapper that abstracts the complexity of different PQC primitives, enabling low-latency execution and high hardware utilization.
    
    \item \textbf{Result Commitment:} The \texttt{committer} stage is responsible for the final handshake via the \texttt{xif-result} interface. It captures the processed data, validates its integrity, and ensures synchronization with the host CPU's pipeline. By managing the write-back of results to the RISC-V general-purpose registers (GPRs), the committer maintains precise exception semantics and strict data consistency. From the perspective of the host processor, a \horcrux-accelerated instruction is completed with the same determinism and transparency as a native arithmetic operation.
\end{itemize}

Adopting CV-X-IF integrates \horcrux into a robust RISC-V ecosystem, providing a standardized, non-invasive method to extend the ISA without modifying \textit{golden} core RTL. 
This \textit{agility-aware} approach implements the standardized bridge recommended by NIST to decouple specialized cryptographic functions from the host CPU. By enabling modular updates that do not require an overhaul of the entire system, \horcrux positions itself as a resilient component in the technology supply chain \cite{nist_cswp_39}.
This interface is now supported by prominent cores and platforms such as CVA6\footnote{\url{https://github.com/openhwgroup/cva6}}, CV32E40X, CV32E40P/PX, CV32E40S, and the X-HEEP\footnote{\url{https://github.com/x-heep/x-heep}} platform. Its versatility is well-proven across AI, signal processing, and security domains~\cite{electronics14153152, 11043433, 10752475, 10.1145/3706594.3726982, s23239408, 10993202}.
To validate its performance within this landscape, \horcrux has been integrated into the open-source \xheep microcontroller platform.
By utilizing instruction-level coupling, the design bypasses the traditional overhead associated with software drivers and memory-mapped registers, enabling seamless portability across the Core-V family without requiring invasive modifications to the host RTL or toolchain.
The final system integration is depicted in \autoref{fig:horcrux}, and the complete implementation is available as open-source code.

\section{Experimental Results}
\label{sec:Results}

This section presents the experimental evaluation of \horcrux, utilizing a SystemVerilog testbench equipped with hardware performance counters for precise, cycle-accurate profiling. \autoref{sec:performances} analyzes the computational speedups achieved by the custom ISA compared to a software \textit{Baseline} (\autoref{table:performance_results_pqc}). Resource utilization and state-of-the-art FPGA comparisons are detailed in \autoref{sec:fppa_results} (\autoref{tab:fpga_comparison}), followed by technology-independent ASIC area metrics in \autoref{sec:asic_results}. Finally, we validate the design's energy efficiency through instruction-level power characterization (\autoref{tab:instr-energy-full}), demonstrating the energy savings gained by offloading critical bottlenecks.

\subsection{Performances}
\label{sec:performances}

To evaluate the computational impact of the proposed extensions, we compare execution times across two distinct configurations: \textit{Baseline}, which represents pure software implementations of the PQC algorithms running on a standard RISC-V CPU, and \textit{Optimized}, which reflects the same routines re-implemented to leverage the \horcrux custom ISA extensions. Both configurations were compiled with the RV32IMACB (\texttt{RV32IMAC\_Zba\_Zbb\_Zbc\_Zbs}) instruction set, ensuring that performance gains are measured against a bit-manip-optimized software baseline.
As shown in \autoref{table:performance_results_pqc}, the acceleration profile varies significantly by algorithmic family. Hash-based schemes (\slhdsa) achieve the highest peak speedup (128.7$\times$) as symmetric permutations map perfectly to the custom \keccak core. Code-based (\hqc) and integer lattice schemes (\mldsa) reach up to 27.1$\times$ and 9.17$\times$ improvements, respectively, benefiting heavily from offloaded vector and modular arithmetic.
Overall, these architectural enhancements extract a substantially larger performance delta than standard assembly optimizations on the Cortex-M4 \cite{cryptoeprint:2019/844}.
\\Security parameter scaling reveals distinct trends. For \mlkem, \mldsa, and \hqc, speedup \textit{increases} with the security level, as larger parameters increase the proportion of time spent in fully offloaded, parallelizable routines. Conversely, \slhdsa speedups \textit{decrease} because exponentially larger tree heights shift the bottleneck from hashing to memory access over the 32-bit bus. 
\\Finally, \falcon yields modest absolute gains (peaking at 2.28$\times$) and an inverse scaling trend due to its reliance on 64-bit floating-point arithmetic. Rather than dedicating massive silicon area to a full FP64 coprocessor, \horcrux adopts an area-efficient strategy where the \texttt{fpr-unit} multiplexes 64-bit operations over the 32-bit RISC-V interface. This creates a structural bottleneck: frequent data shuffling leaves \falcon's performance bound by bus latency and Amdahl's Law, a constraint magnified at higher security levels (e.g., \falcon-1024) which demand even more intensive data multiplexing.

\begin{table}[h]
\centering
\caption{Cycle Count Comparison: \textit{Baseline} vs. \textit{Optimized} Implementations.}
\label{table:performance_results_pqc}
\scriptsize
\setlength{\tabcolsep}{3pt}
\renewcommand{\arraystretch}{1.2}

\resizebox{\columnwidth}{!}{%
\begin{tabular}{l|c|cc|cc|cc}
\hline
\multirow{2}{*}{\textbf{Algorithm}} &
\multirow{2}{*}{\textbf{Ver.}} & \multicolumn{2}{c}{\textbf{KeyGen}} &
\multicolumn{2}{c}{\textbf{Enc/Sign}} &
\multicolumn{2}{c}{\textbf{Dec/Ver}} \\
\cline{3-8}
& & kCC & SU & kCC & SU & kCC & SU \\
\hline\hline

\rowcolor{gray!10}
\multicolumn{8}{l}{\textit{Lattice-based KEM (\mlkem)}} \\
\hline
\multirow{2}{*}{\mlkem-512}
& B & 900 & -- & 965 & -- & 1,108 & -- \\
& O & 163 & \speedup{5.52} & 189 & \speedup{5.11} & 290 & \speedup{3.82} \\
\hline
\multirow{2}{*}{\mlkem-768}
& B & 1,475 & -- & 1,602 & -- & 1,789 & -- \\
& O & 260 & \speedup{5.67} & 289 & \speedup{5.54} & 426 & \speedup{4.20} \\
\hline
\multirow{2}{*}{\mlkem-1024}
& B & 2,357 & -- & 2,500 & -- & 2,739 & -- \\
& O & 389 & \speedup{6.06} & 421 & \speedup{5.94} & 601 & \speedup{4.56} \\
\hline

\rowcolor{gray!10}
\multicolumn{8}{l}{\textit{Code-based KEM (\hqc)}} \\
\hline
\multirow{2}{*}{\hqc-1}
& B & 46,108 & -- & 92,006 & -- & 140,074 & -- \\
& O & 2,316 & \speedup{19.9} & 4,357 & \speedup{21.1} & 7,123 & \speedup{19.7} \\
\hline
\multirow{2}{*}{\hqc-3}
& B & 139,965 & -- & 279,352 & -- & 421,874 & -- \\
& O & 6,801 & \speedup{20.6} & 12,967 & \speedup{21.5} & 20,039 & \speedup{21.1} \\
\hline
\multirow{2}{*}{\hqc-5}
& B & 344,053 & -- & 687,260 & -- & 1,024,298 & -- \\
& O & 13,323 & \speedup{25.8} & 25,334 & \speedup{27.1} & 38,695 & \speedup{24.5} \\
\hline

\rowcolor{gray!10}
\multicolumn{8}{l}{\textit{Lattice-based Signature (\mldsa)}} \\
\hline
\multirow{2}{*}{\mldsa-44}
& B & 3,444 & -- & 6,820 & -- & 3,618 & -- \\
& O & 458 & \speedup{7.52} & 1,537 & \speedup{4.44} & 592 & \speedup{6.11} \\
\hline
\multirow{2}{*}{\mldsa-65}
& B & 6,041 & -- & 23,455 & -- & 6,084 & -- \\
& O & 790 & \speedup{7.65} & 6,177 & \speedup{3.80} & 913 & \speedup{6.66} \\
\hline
\multirow{2}{*}{\mldsa-87}
& B & 10,235 & -- & 26,738 & -- & 10,394 & -- \\
& O & 1,116 & \speedup{9.17} & 6,360 & \speedup{4.20} & 1,386 & \speedup{7.50} \\
\hline

\rowcolor{gray!10}
\multicolumn{8}{l}{\textit{Stateless Hash-based Signature (\slhdsa)}} \\
\hline
\multirow{2}{*}{\slhdsa-128-fs}
& B & 138,814 & -- & 1,627,526 & -- & 190,001 & -- \\
& O & 1,905 & \speedup{72.9} & 45,335 & \speedup{35.9} & 2,930 & \speedup{64.8} \\
\hline
\multirow{2}{*}{\slhdsa-128-fr}
& B & 262,561 & -- & 1,835,864 & -- & 375,824 & -- \\
& O & 2,041 & \speedup{128.7} & 48,582 & \speedup{37.8} & 3,223 & \speedup{116.6} \\
\hline
\multirow{2}{*}{\slhdsa-192-fs}
& B & 201,918 & -- & 991,317 & -- & 275,172 & -- \\
& O & 3,553 & \speedup{56.8} & 96,253 & \speedup{10.3} & 5,445 & \speedup{50.5} \\
\hline
\multirow{2}{*}{\slhdsa-192-fr}
& B & 384,731 & -- & 1,283,162 & -- & 555,947 & -- \\
& O & 3,729 & \speedup{103.2} & 101,150 & \speedup{12.7} & 5,956 & \speedup{93.3} \\
\hline
\multirow{2}{*}{\slhdsa-256-fs}
& B & 536,735 & -- & 2,416,664 & -- & 293,555 & -- \\
& O & 11,258 & \speedup{47.7} & 240,821 & \speedup{10} & 6,902 & \speedup{42.5} \\
\hline
\multirow{2}{*}{\slhdsa-256-fr}
& B & 1,024,313 & -- & 3,325,795 & -- & 558,107 & -- \\
& O & 12,064 & \speedup{84.9} & 249,327 & \speedup{13.3} & 7,254 & \speedup{76.9} \\
\hline

\rowcolor{gray!10}
\multicolumn{8}{l}{\textit{Lattice-based Signature (\falcon)}} \\
\hline
\multirow{2}{*}{\falcon-512}
& B & 162,348 & -- & 67,331 & -- & 585 & -- \\
& O & 117,786 & \speedup{1.38} & 48,610 & \speedup{1.39} & 258 & \speedup{2.27} \\
\hline
\multirow{2}{*}{\falcon-1024}
& B & 576,874 & -- & 147,135 & -- & 1,207 & -- \\
& O & 437,859 & \speedup{1.32} & 116,874 & \speedup{1.26} & 530 & \speedup{2.28} \\
\hline

\multicolumn{8}{l}{\textit{B = Baseline, O = Optimized, kCC = kCycles, SU = Speedup factor}}\\
\multicolumn{8}{l}{\textit{\textbf{Note}: Cycle counts represent averages over 100 KAT executions.}}
\end{tabular}
}
\end{table}

\subsection{FPGA Results}
\label{sec:fppa_results}

\horcrux resource utilization was assessed on the AMD Zynq UltraScale+ ZCU104 using \textbf{Vivado 2022.2}. \autoref{tab:fpga_utilization} details the area overhead of the extended system, while \autoref{tab:fpga_comparison} contrasts our implementation with state-of-the-art FPGA accelerators, demonstrating \horcrux maintains a competitive footprint alongside broader algorithm coverage.

Before benchmarking against dedicated hardware accelerators, we situate \horcrux against optimized software implementations using select cryptographic kernels to illustrate the practical trade-offs of our tightly coupled ISE. 
For highly symmetric workloads, the benefits of direct hardware offloading are clear: \horcrux executes a full \keccak permutation in 385 clock cycles, achieving a $\sim$20$\times$ reduction over the 7,808 cycles required by optimized single-issue 32-bit software \cite{cryptoeprint:2024/1515}. For lattice arithmetic, our hardware NTT and INTT require 14,755 and 17,754 cycles, respectively. While this improves upon standard non-SIMD software baselines, 
it inherently trails the performance of aggressive superscalar dual-issue software implementations \cite{cryptoeprint:2024/1515} (5,714 and 6,005 cycles). Rather than asserting absolute performance superiority across all software paradigms, these representative examples highlight \horcrux's primary architectural intent: delivering competitive acceleration bounded within a standard single-issue pipeline, thereby bypassing the severe area and power overheads intrinsic to complex dual-issue CPU architectures.
Ultimately, these kernel-level hardware efficiencies translate into dominant protocol-level performance. For \mlkem-768, \horcrux completes key generation, encapsulation, and decapsulation in 260k, 289k, and 426k cycles—roughly twice as fast as the state-of-the-art software equivalents (496k, 606k, and 578k cycles) reported in \cite{cryptoeprint:2024/1515}. Similarly, \horcrux delivers substantial improvements over their \mldsa-44 software baseline, achieving up to a 9$\times$ end-to-end speedup. These results validate our hardware-software co-design approach, proving that strategic ISEs offer a vastly superior performance-to-area tradeoff compared to pure assembly optimizations.

\begin{table}[h]
\centering
\caption{Resource Utilization FPGA (AMD Zynq UltraScale+ ZCU104).}
\label{tab:fpga_utilization}
\scriptsize
\setlength{\tabcolsep}{4pt} 
\renewcommand{\arraystretch}{1.1}

\resizebox{\columnwidth}{!}{
\begin{tabular}{l c c c c }
\hline
\rowcolor{gray!10} \textbf{Architecture} & \textbf{LUTs} & \textbf{Registers} & \textbf{BRAMs} & \textbf{DSPs} \\ \hline \hline
\xheep (Original) & 71,630 & 32,750 & 256 & 9 \\ \hline 
\textbf{\xheep } & \textbf{91,826} & \textbf{37,179} & \textbf{256} & \textbf{20} \\ 
\textbf{+ \horcrux} & \textcolor{speedupblue}{\textbf{[+20,196]}} & \textcolor{speedupblue}{\textbf{[+4,429]}} & \textcolor{speedupblue}{\textbf{[+0]}} & \textcolor{speedupblue}{\textbf{[+11]}} \\ 
\hspace{0.5em} $\diamond$ \texttt{decoder} & 5,611 & 164 & 0 & 0 \\ 
\hspace{0.5em} $\diamond$ \texttt{PQ-ALU} & 14,582 & 4,260 & 0 & 11 \\ 
\hspace{1em} $\diamond$ \texttt{horcrux\_register} & 3,873 & 1,600 & 0 & 0 \\ 
\hspace{0.5em} $\diamond$ \texttt{committer} & 5 & 0 & 0 & 0 \\ \hline
\end{tabular}%
} 
\end{table}

The \horcrux platform introduces a targeted area overhead, primarily due to cryptographic primitives and synchronization logic (\autoref{tab:fpga_utilization}). The \keccak unit accounts for 6,995 LUTs and 1,615 FFs, representing approximately 35\% of the total \horcrux LUT overhead; this footprint is necessary to manage the complete internal state and the intensive logic of the \keccak permutation. Furthermore, the \texttt{horcrux\_register} serves as the central data hub, utilizing 3,873 LUTs for control logic and 1,600 FFs as dedicated memory for the intermediate state, ensuring efficient coordination across the modules.

\textbf{Comparisons.} \autoref{tab:fpga_comparison} compares our work against prior FPGA implementations. Most existing designs target only one or, at most, two schemes, whereas \horcrux supports all listed schemes within a single unified architecture. Our design on ZCU104 requires 5,049 eSlices\footnote{The metric eSlices is computed as $\max\{\text{LUT}/4,\ \text{FF}/8\}$.}, which is competitive with, and in several cases smaller than, specialized single-scheme implementations. 
In contrast, state-of-the-art designs incur significantly higher resource usage when considered collectively, highlighting the efficiency of our crypto-agile approach. Even under an optimistic assumption where the smallest reported implementations for each scheme are combined, i.e., \cite{9605604} (\mlkem/\mldsa), \cite{11414189} (\hqc), \cite{10294284} (\mlkem), \cite{cryptoeprint:2023/1505} (\mldsa), \cite{Sphincs_hashes} (\slhdsa), and \cite{10.1145/3579092} (\falcon), the total area amounts to 10,072 eSlices, which is more than twice that of our design, corresponding to an area reduction of approximately 51\,\%, and further demonstrating the effectiveness of \horcrux in achieving compact multi-scheme support without proportional overhead. 
Crucially, while integrating the unified \texttt{multiplier-tree} extends the critical path and caps the frequency at 42 MHz (compared to 125 MHz for \horcrux without this unit), this is a deliberate architectural trade-off. Deeply constrained edge devices inherently operate at lower clock speeds to satisfy strict energy budgets. By intentionally trading peak frequency for maximum hardware reuse, \horcrux prioritizes what matters most in this domain: delivering comprehensive cryptographic agility while remaining strictly within tight area and power limits.

\begin{table}[h]
\centering
\caption{Comparison with State-of-the-Art FPGA Implementations.}
\label{tab:fpga_comparison}
\scriptsize
\setlength{\tabcolsep}{4pt} 
\renewcommand{\arraystretch}{1.2}

\resizebox{\columnwidth}{!}{
\begin{tabular}{l c r r r r r r}
\toprule
\multirow{2}{*}{\textbf{Ref.}} & \multirow{2}{*}{\textbf{Device}} & \multicolumn{4}{c}{\textbf{Resources}} & \multirow{2}{*}{\textbf{eSlices}} & \textbf{\textit{$f_{max}$}} \\
\cmidrule(lr){3-6}
& & \textbf{LUT} & \textbf{FF} & \textbf{DSP} & \textbf{BRAM} & & [MHz] \\
\hline

\rowcolor{gray!10} \multicolumn{8}{l}{\textit{Lattice-based KEM / Signature (\mlkem, \mldsa, \falcon)}} \\\hline
\cite{9946370}$^{\blacktriangle}$ & ZCU102 & 23,347 & 9,798 & 4 & 24 & 5,837 & 270 \\
\cite{cryptoeprint:2024/1192}$^{\diamond}$ & Kintex-7 & 48,158 & 16,641 & 112 & 0 & 12,040 & - \\
\cite{Ye_Song_Zhang_Chen_Cheung_Huang_2024} & Zynq-7000 & 24,177 & 8,859 & 37 & 24 & 6,044 & 125 \\
\cite{9605604} & ZCU106 & 22,451 & 10,721 & 10 & 0.5 & 5,613 & 100 \\
\cite{10.1007/978-981-95-8399-7_12} & ZCU102 & 37,890 & 15,753 & 20 & 15 & 9,473 & 200 \\
\cite{PQC_OpenTitan} & Kintex-7 & 6,337 & 1,006 & 33 & 5 & 1,584 & 100 \\
\cite{cryptoeprint:2023/1505} & Zynq-7000 & 360 & 16 & 2 & 0 & 90 & 20 \\
\cite{10.1145/3643826} & Zynq-7000 & 3,994 & 1,605 & 7 & 0 & 999 & 100 \\
\cite{10.1145/3579092} & ZCU102 & 7,219 & 3,238 & 7 & 6 & 1,805 & -- \\
\cite{9609917}$^{\blacksquare}$ & Artix-7 & 53,187 & 28,318 & 16 & 29 & 13,297 & 116 \\
\cite{cryptoeprint:2023/1885}$^{\blacksquare}$ & ZCU104 & 13,302 & 8,619 & 15 & 14 & 3,326 & 214 \\
\cite{10294284} & Artix-7 & 93 & 0 & 1 & 0 & 23 & 32.9 \\
\cite{cryptoeprint:2020/049}$^{\diamond}$ & Artix-7 & 104 & 35 & 1 & 0 & 26 & 59.2 \\
\cite{Fritzmann_Sigl_Sepúlveda_2020} & Zynq-7000 & 9,058 & 1,268 & 0 & 12 & 2,265 & -- \\

\hline
\rowcolor{gray!10} \multicolumn{8}{l}{\textit{Code-based KEM (\hqc)}} \\\hline
\cite{10993202} & Kintex-7 & 8,894 & 3,710 & 0 & 0 & 2,224 & 150 \\
\cite{11414189} & Artix-7 & 4,550 & 1,757 & 0 & 0 & 1,138 & - \\
\cite{cryptoeprint:2025/601}$^{\blacktriangle}$ & Zynq-7020 & 5,076 & 4,083 & 0 & 8.5 & 1,269 & 125 \\

\hline
\rowcolor{gray!10} \multicolumn{8}{l}{\textit{Hash-based Signature (\slhdsa)}} \\\hline
\cite{9217834}$^{\blacksquare}$ & Artix-7 & 51,009 & 74,539 & 1 & 22.5 & 12,752 & 250 \\
\cite{9460703}$^{\blacksquare}$ & XZCU3EG & 7,925 & 6,339 & 0 & 0.5 & 1,981 & 152 \\
\cite{Sphincs_hashes} & Spartan-7 & 5,611 & 2,652 & 0 & 32 & 1,403 & 100 \\

\hline\hline
\textbf{This Work} & \textbf{ZCU104} & \textbf{20,196} & \textbf{4,429} & \textbf{11} & \textbf{0} & \textbf{5,049} & \textbf{42} \\
\bottomrule
\multicolumn{8}{l}{$^{\diamond}$ Vectorized implementations. \quad $^{\blacksquare}$ Standalone implementations.} \\
\multicolumn{8}{l}{$^{\blacktriangle}$ Loosely-coupled implementations.} \\
\end{tabular}
} 
\end{table}

\subsection{ASIC Results}
\label{sec:asic_results}

Synthesized using Synopsys Design Compiler (65nm CMOS) at 160 MHz, \horcrux occupies \textbf{115.8 kGE}. \autoref{tab:instr-energy-full} details instruction-level energy (nJ). By minimizing cycles and memory transfers via internal state management, \horcrux significantly reduces total energy per operation versus software. Finally, we benchmark ATP and overall efficiency against state-of-the-art ASIC implementations.

\textbf{Power and Energy measurements.} We characterize power to demonstrate that our architectural enhancements significantly reduce the overall energy footprint. \autoref{tab:instr-energy-full} compares the instruction-level energy of software baselines against \horcrux. The evaluated test suite maps exactly to the parenthetical bottleneck names defined in \autoref{sec:Profiling}.

\begin{table*}[h]
\centering
\caption{Instruction-level characterization. Legend: \textit{s} = software, \textit{c} = custom. Energy reported in Nano-Joules [nJ].}
\label{tab:instr-energy-full}
\begin{tabular}{l >{\columncolor{gray!10}}c c >{\columncolor{gray!10}}c c >{\columncolor{gray!10}}c c >{\columncolor{gray!10}}c c >{\columncolor{gray!10}}c c c r}
\toprule
\textbf{Test} & \textbf{C.C.$_s$} & \textbf{C.C.$_c$} & \textbf{P$_{tot,s}$} & \textbf{P$_{tot,c}$} & \textbf{P$_{sw,s}$} & \textbf{P$_{sw,c}$} & \textbf{P$_{int,s}$} & \textbf{P$_{int,c}$} & \textbf{E$_s$} & \textbf{E$_c$} & \textbf{$\delta E$} & \textbf{E. Sav.} \\
 & [cycles] & [cycles] & [\textmu W] & [\textmu W] & [\textmu W] & [\textmu W] & [\textmu W] & [\textmu W] & [nJ] & [nJ] & [nJ] & $\times$ Spd \\
\midrule
sha3\_256sb & 24,548 & 1,637 & 183 & 250 & 9.5 & 32.2 & 21.7 & 65 & 28.1 & 2.6 & 25.5 & \textbf{$\times$11} \\
sha3\_256mb & 50,253 & 3,132 & 183 & 293 & 9.5 & 47 & 21.8 & 91 & 57.5 & 5.7 & 51.7 & \textbf{$\times$10} \\
shake256    & 27,286 & 3,260 & 183 & 233 & 9.4 & 26.9 & 21.2 & 53.7 & 31.2 & 4.7 & 26.5 & \textbf{$\times$6.6} \\
multi\_squeeze & 78,580 & 4,796 & 183 & 276 & 9.4 & 40.6 & 21.7 & 81.5 & 89.9 & 8.3 & 81.6 & \textbf{$\times$11} \\
gen-matrix  & 418,060 & 24,763 & 183 & 777 & 9.4 & 208 & 21.6 & 416 & 478.2 & 120.3 & 357.9 & \textbf{$\times$4} \\

\midrule
prf-addr    & 24,540 & 382 & 183 & 525 & 9.42 & 127 & 21.3 & 245 & 28.1 & 1.25 & 26.8 & \textbf{$\times$22.4} \\
chain-len (128)\textsuperscript{$\ddagger$} & 552 & 160 & 187 & 438 & 9.7 & 118 & 25.6 & 167 & 0.65 & 0.34 & 0.30 & \textbf{$\times$1.5} \\
thash       & 49,666 & 408 & 183 & 760 & 9.43 & 203 & 21.6 & 403 & 56.8 & 1.94 & 54.87 & \textbf{$\times$29.3} \\
gen\_chain  & 745,008 & 6,183 & 183 & 777 & 9.4 & 208 & 21.6 & 416 & 852.1 & 30 & 822.1 & \textbf{$\times$28.4} \\
compute\_root & 151,750 & 1,444 & 183 & 712 & 9.42 & 186 & 21.5 & 371 & 173.6 & 6.4 & 127.1 &  \textbf{$\times$27.} \\

\midrule
mlkem-poly-intt  & 60,923 & 21,861 & 326 & 789 & 36.8 & 191 & 139 & 448 & 124.1 & 107.8 & 16.32 & \textbf{$\times$1.2} \\
mlkem-poly-ntt   & 45,878 & 24,819 & 326 & 590 & 36.9 & 162 & 139 & 330 & 93.5 & 91.5 & 1.96 & \textbf{$\times$1.02} \\
mldsa-poly-ntt   & 191,854 & 62,853 & 325	& 709 & 36.3 &	177 &	139	& 382 & 389.7 &	278.5 &	111.2 & \textbf{$\times$1.4} \\
mldsa-poly-intt   &  221,326	& 74,998	& 325 & 682	& 36.3	& 165	& 139 & 366	& 449.6 &	319.7 & 129.9 &  \textbf{$\times$1.4} \\
falcon-intt & 2,085 & 1,209 & 170 & 259 & 7.71 & 41.1 & 10.1 & 65.5 & 2.21 & 1.96 & 0.26 & \textbf{$\times$1.1} \\
falcon-ntt  & 1,653 & 996 & 171 & 257 & 7.84 & 39.3 & 10.9 & 65.4 & 1.77 & 1.60 & 0.17 & \textbf{$\times$1.1} \\
falcon-fpr-norm & 10,142 & 1,913 & 163 & 296 & 6.86 & 51.7 & 3.91 & 92 & 10.3 & 3.5 & 6.79& \textbf{$\times$2.9} \\
f-mq-montymul\textsuperscript{$\ddagger$} & 359 & 113 & 172 & 353 & 7.99 & 82.8 & 12.2 & 118 & 0.39 & 0.25 & 0.14 & \textbf{$\times$1.6} \\
gf-carryless\textsuperscript{$\ddagger$} & 3,047 & 114 & 174 & 271 & 8.2 & 53.6 & 13.4 & 65 & 3.3 & 0.19 & 3.12 & \textbf{$\times$17.2} \\
karats     & 2,634 & 8 & 172 & 374 & 7.93 & 88.5 & 11.7 & 133 & 2.8 & 0.02 & 2.81 & \textbf{$\times$151} \\
gf-reduce\textsuperscript{$\ddagger$} & 256 & 104 & 174 & 201 & 8.16 & 24.3 & 13.4 & 24.5 & 0.28 & 0.13 & 0.15 & \textbf{$\times$2.1} \\

\midrule
cbd\_eta2\textsuperscript{$\dagger$} & 11,031 & 3,835 & 177 & 208 & 8.46 & 24.3 & 16.3 & 32 & 0.12 & 0.05 & 0.07 & \textbf{$\times$2.5} \\
cbd\_eta3\textsuperscript{$\dagger$} & 7,123 & 2,036 & 177 & 208 & 8.46 & 24.3 & 16.3 & 32 & 0.079 & 0.026 & 0.05 & \textbf{$\times$3} \\
mldsa-rej-eta\textsuperscript{$\dagger$} & 1,684 & 400 & 174 & 183 & 8.24 & 12.1 & 14.2 & 18.8 & 1.83 & 0.46 & 1.38 & \textbf{$\times$4} \\
\bottomrule
\multicolumn{13}{l}{\textsuperscript{$\dagger$} Test performed with \textit{N=100}, \textsuperscript{$\ddagger$} Test performed with \textit{N=10}} \\

\end{tabular}

\end{table*}

For each function, we report the clock cycles ($C.C.$) together with the total ($P_{tot}$), switching ($P_{sw}$), and internal ($P_{int}$) power consumption. These values are obtained from post-synthesis netlist simulations, with switching activity analyzed using \textbf{Synopsys PrimePower 2020.09}. Energy consumption is computed at an operating frequency of 160~MHz as $E = (P_{tot} \times C.C. / f_{req}) / N$, where $N$ denotes the number of executed tests. The results are reported both in absolute terms ($E$) and as the difference between the software baseline and the custom instruction ($\delta E$). Finally, the table summarizes the overall efficiency gains in terms of energy speedup, defined as $E_s / E_c$.
The results show massive efficiency gains—notably for \texttt{karats} (151$\times$) and \texttt{thash} (29.3$\times$), primarily driven by the substantial reduction in clock cycles enabled by the proposed custom hardware.
Conversely, the custom implementations exhibit higher total power consumption than the software baseline, largely due to the absence of power-oriented optimizations such as clock gating and operand isolation at this design stage.
Furthermore, the computationally dense \mlkem and \mldsa polynomial operations (e.g., wide-datapath NTT butterflies) induce high concurrent switching activity, resulting in lower relative energy speedups and elevated power profiles.
Nevertheless, across the full set of evaluated primitives, the significant reduction in execution time consistently translates into an overall energy advantage, confirming the effectiveness of \horcrux.

\textbf{Comparisons.} \autoref{tab:ASIC_comparison} benchmarks \horcrux against prior ASIC implementations at the highest security levels. While specialized accelerators naturally yield superior ATP for specific algorithms, \horcrux distinguishes itself through unmatched crypto-agility. Most existing works target only one or two schemes, and ASIC literature for \hqc and \falcon remains exceptionally sparse. The only comparable multi-algorithm ASIC design \cite{s23239408} (54 kGE in 28 nm) reports merely function-level metrics, while \horcrux provides complete end-to-end execution and energy validation across all five NIST PQC families, all while maintaining a highly competitive unified footprint of 115.8 kGE.

\begin{table*}[h]
\centering
\caption{Comparison with State-of-the-Art ASIC Implementations. Area and Area-Time Product (ATP) results are reported as kGE / kCellCount (kC).}
\label{tab:ASIC_comparison}
\resizebox{\textwidth}{!}{
\begin{tabular}{c c c c c ccc}
\hline
\multirow{2}{*}{\textbf{Ref.}} & \multirow{2}{*}{\textbf{Scheme}} & \textbf{Tech.} & \textbf{Area} & \textbf{\textit{f}} &
\multicolumn{3}{c}{\textbf{Area-Time Product (ATP)}} \\
& & \textbf{[nm]} & \textbf{[kGE / kC]} & \textbf{[MHz]} & \textbf{K} & \textbf{E/S} & \textbf{D/V} \\
\hline

\rowcolor{gray!10} \multicolumn{8}{l}{\textit{Lattice-based KEM / Signature (\mlkem, \mldsa, \falcon)}} \\ \hline
\multirow{2}{*}{\cite{9946370} \textsuperscript{$\blacktriangle$}} & ML-KEM-1024 & \multirow{2}{*}{65} & 769 / - & \multirow{2}{*}{280} & 25,915 / - & 32,328 / - & 38,189 / - \\
& ML-DSA-87 & & 769 / - & & 109,134 / - & 192,742 / - & 128,179 / - \\ \hline

\multirow{2}{*}{\cite{cryptoeprint:2024/1192}} \textsuperscript{$\diamond$} & \mlkem-1024 & \multirow{2}{*}{7} & \multirow{2}{*}{- / 120} & \multirow{2}{*}{100} & - /122,168 & - / 135,652 & - / 153,187 \\
& \mldsa-65  &  &  &  & - / 438,601 & - / 788,714 & - / 434,058 \\ \hline


\cite{Fritzmann_Sigl_Sepúlveda_2020} & ML-KEM-1024 & 65 & - / 21.2 & 45 & - / 164,889 & - / 190,800 & - / 200,222 \\ \hline

\cite{10562296} & ML-KEM-1024 & 28 & - / 9.38 & 500 & - / 65,681 & - / 80,694 & - / 80,694 \\ \hline

\multirow{2}{*}{\cite{10.1145/3579092}}$^{\blacktriangle}$ & ML-DSA-87 & \multirow{2}{*}{22} & - / 32.99 & \multirow{2}{*}{800} & - / - & - / - & - / 336,086 \\
& FALCON-1024 & & - / 32.99 & & - / - & - / - & - / 25,320 \\ \hline

\cite{10294284} & ML-KEM-1024 & 55 & 0.799 / - & 32.9 & 37,473 / - & 44,953 / - & 41,747 / - \\ \hline

\cite{10.1145/3643826} & ML-DSA-87 & 55 & - / 12.4 & 100 & - / 190,121 & - / 256,117 & - / 193,567 \\ \hline

\cite{cryptoeprint:2023/1505} & ML-DSA-65 & 28 & - / 3.16 & 100 & - / 125,452 & - / 133,289 & - / 117,299 \\ \hline

\cite{PQC_OpenTitan} & \mldsa-87 & 22 & 242 & 100 & - & - & 2,892,490 \\ \hline

\multirow{2}{*}{\cite{Ye_Song_Zhang_Chen_Cheung_Huang_2024}} 
& \mlkem-1024 & \multirow{2}{*}{28} & \multirow{2}{*}{149.1 / -} & \multirow{2}{*}{200} & 145,017 / - & 147,466 / -	& 182,027 / -\\
& \mldsa-65 & & & & 348,486	/ - & 874,114 / - & 399,993 / -\\ \hline

\rowcolor{gray!10} \multicolumn{8}{l}{\textit{Hash-based Signature (\slhdsa)}} \\ \hline
\cite{10915711} & \slhdsa-256fs & 28 & 115.2 / -   & 200 & 13,701,888/ -   & 282,995,712 / -  & 7,236,288 / -   \\ 

\hline\hline
\multirow{5}{*}{\textbf{This Work}} 
& \mlkem-1024 & \multirow{5}{*}{\textbf{65}} & \multirow{5}{*}{\textbf{115.8 / 65}} & \multirow{5}{*}{\textbf{160}} 

& 281,538 / 158,031 & 304,698 / 171,031 & 434,973 / 244,156\\

& \hqc-5 & & & &  9,642,429 / 5,412,417 &	18,335,787 / 10,292,108	&	28,005,212 / 15,719,678 \\

& \mldsa-87 & & & & 807,705 / 453,375 & 4,603,050 / 2,583,750 & 1,003,117 / 563,062 \\

& \slhdsa-256fs & & & &  8,148,322/4,573,755 & 174,294,314/97,833,596 & 4,995,009/2,803,762 \\

& \falcon-1024 & & & & 316,900,549 / 177,880,274  & 84,587,342 / 47,479,941	&	383,827 /215,447\\ \hline

\multicolumn{8}{l}{$^{\diamond}$ Vectorized implementations. \quad $^{\blacksquare}$ Standalone implementations. \quad $^{\blacktriangle}$ Loosely-coupled implementations.}

\end{tabular}
}
\end{table*}

\section{Discussion and Future Works}
\label{sec:discussion}

A guiding principle throughout the development of \horcrux has been crypto-agility. Rather than treating algorithmic resource needs in isolation, we implement the cross-paradigm optimization highlighted by NIST as a critical new \textit{area of research} \cite{nist_cswp_39}. By unifying the datapath for lattice-based (\mlkem, \mldsa, \falcon), hash-based (\slhdsa), and code-based (\hqc) schemes around shared versatile primitives, \horcrux directly addresses the ``\textit{Resource Considerations}'' that NIST identifies as the primary hurdle for hardware agility \cite{nist_cswp_39}. This functional synergy avoids rigid, monolithic accelerators and ensures future-proof adaptability. For instance, the hardware can accommodate new key sizes or pseudo-Mersenne primes simply by updating decoder constants, bypassing the need for structural overhauls.
Crucially, this design directly addresses the severe trade-offs inherent in deeply constrained environments. Our primary target use case comprises battery-powered embedded systems characterized by limited silicon area, low operating frequencies, and strict energy budgets. By sacrificing a high-frequency, heavily pipelined critical path in favor of a highly dense, unified arithmetic unit, \horcrux delivers essential PQC acceleration while remaining comfortably within strict power and area limits. This ensures devices can safely transition to PQC and maintain flexibility via fallback mechanisms without incurring prohibitive overhead.

Regarding modular integration, leveraging the CV-X-IF interface provides the standardized API necessary to decouple cryptographic extensions from core application logic \cite{nist_cswp_39}. This agility-aware strategy ensures \horcrux acts as a resilient component in the technology supply chain, allowing for future algorithm replacements without altering the golden core RTL. While validated alongside the CV32E40PX, the architectural benefits of instruction-level offloading and internal state persistence are fundamentally interface-agnostic and easily portable to other RISC-V cores.

From a security perspective, \horcrux is natively protected against timing side-channel attacks through strict fixed-latency instruction execution. To complete the security profile, future work will prioritize Side-Channel Analysis (SCA) countermeasures, specifically hardening the architecture against power analysis. The modular nature of our datapath inherently supports this evolution; functional units and the internal register file can be replaced with masked variants without disrupting the host CPU's pipeline. 

Ultimately, by balancing low area and energy consumption with comprehensive multi-scheme support, \horcrux establishes an effective, future-proof foundation for crypto-agile RISC-V embedded systems.

\section{Conclusion}
\label{sec:conclusion}

This work presents \horcrux, a RISC-V Instruction Set Extension supporting a comprehensive PQC suite, \mlkem, \mldsa, \slhdsa, \hqc, and \falcon, within a single unified coprocessor. Moving beyond isolated, single-scheme accelerators, \horcrux offloads core cryptographic bottlenecks (such as \keccak, modular/Galois arithmetic, and sampling) while maintaining full compatibility with the standard RISC-V toolchain. Experimental results demonstrate massive execution speedups,up to 129$\times$ for hash-based, 27$\times$ for code-based, and 9.17$\times$ for lattice-based schemes, all while achieving a smaller resource footprint than many state-of-the-art dedicated architectures. Crucially, by replacing algorithm-specific logic with versatile, shared hardware like the unified multiplier tree, \horcrux prioritizes crypto-agility and maximal resource reuse. Ultimately, this approach provides a highly area-efficient, future-proof reference for NIST-compliant HW/SW co-design tailored to constrained embedded systems.

\bibliographystyle{ieeetr}
\bibliography{mybib}

@techreport{nistPQCReport,
  author      = {Dustin Moody and Ray Perlner and Daniel Smith-Tone and Angela Robinson and Rene Peralta and Lily Chen},
  title       = {{Post-Quantum Cryptography: NIST’s Plan for the Future}},
  institution = {National Institute of Standards and Technology},
  year        = {2016},
  number      = {NISTIR 8105},
  url         = {https://doi.org/10.6028/NIST.IR.8105}
}

@ARTICLE{10562296,
author={Gewehr, Carlos and Luza, Lucas and Moraes, Fernando Gehm},
journal={IEEE Access},
title={{Hardware Acceleration of Crystals-Kyber in Low-Complexity Embedded Systems With RISC-V Instruction Set Extensions}},
year={2024},
volume={12},
number={},
pages={94477-94495},
doi={10.1109/ACCESS.2024.3416812}}

@article{Fritzmann_Sigl_Sepúlveda_2020, title={{RISQ-V: Tightly Coupled RISC-V Accelerators for Post-Quantum Cryptography}}, volume={2020},  url={https://tches.iacr.org/index.php/TCHES/article/view/8683},  DOI={10.13154/tches.v2020.i4.239-280}, number={4}, journal={IACR Transactions on Cryptographic Hardware and Embedded Systems}, author={Fritzmann, Tim and Sigl, Georg and Sepúlveda, Johanna}, year={2020}, month={Aug.}, pages={239–280} }

@article{Banerjee_Ukyab_Chandrakasan_2019, title={{Sapphire: A Configurable Crypto-Processor for Post-Quantum Lattice-based Protocols}}, volume={2019},  url={https://tches.iacr.org/index.php/TCHES/article/view/8344},  DOI={10.13154/tches.v2019.i4.17-61}, number={4}, journal={IACR Transactions on Cryptographic Hardware and Embedded Systems}, author={Banerjee, Utsav and Ukyab, Tenzin S. and Chandrakasan, Anantha P.}, year={2019}, month={Aug.}, pages={17–61} }

@ARTICLE{9605604,
author={Nannipieri, Pietro and Di Matteo, Stefano and Zulberti, Luca and Albicocchi, Francesco and Saponara, Sergio and Fanucci, Luca},
journal={IEEE Access},
title={{A RISC-V Post Quantum Cryptography Instruction Set Extension for Number Theoretic Transform to Speed-Up CRYSTALS Algorithms}},
year={2021},
volume={9},
number={},
pages={150798-150808},
doi={10.1109/ACCESS.2021.3126208}}

@INPROCEEDINGS{8715173,
  author={Fritzmann, Tim and Sharif, Uzair and Müller-Gritschneder, Daniel and Reinbrecht, Cezar and Schlichtmann, Ulf and Sepulveda, Johanna},
  booktitle={2019 Design, Automation \& Test in Europe Conference \& Exhibition (DATE)}, 
  title={{Towards Reliable and Secure Post-Quantum Co-Processors based on RISC-V}}, 
  year={2019},
  volume={},
  number={},
  pages={1148-1153},
  doi={10.23919/DATE.2019.8715173}}

@ARTICLE{10915711,
  author={Ye, Zewen and Li, Xin and Wang, Chuhui and Cheung, Ray C. C. and Huang, Kejie},
  journal={IEEE Transactions on Very Large Scale Integration (VLSI) Systems}, 
  title={{RVSLH: Acceleration of Postquantum Standard SLH-DSA With Customized RISC-V Processor}}, 
  year={2025},
  volume={},
  number={},
  pages={1-5},
  doi={10.1109/TVLSI.2025.3543352}}

@article{Ye_Song_Zhang_Chen_Cheung_Huang_2024, title={{A Highly-efficient Lattice-based Post-Quantum Cryptography Processor for IoT Applications}}, volume={2024},  url={https://tches.iacr.org/index.php/TCHES/article/view/11423},  DOI={10.46586/tches.v2024.i2.130-153}, number={2}, journal={IACR Transactions on Cryptographic Hardware and Embedded Systems}, author={Ye, Zewen and Song, Ruibing and Zhang, Hao and Chen, Donglong and Cheung, Ray Chak-Chung and Huang, Kejie}, year={2024}, month={Mar.}, pages={130–153} }

@misc{nist-pqc-faq,
  title        = {{Post-Quantum Cryptography — FAQs}},
  author       = {{NIST}},
  howpublished = {\url{https://csrc.nist.gov/projects/post-quantum-cryptography/faqs}},
  note         = {Accessed: 2025-05-09},
  year         = {2025}
}

@INPROCEEDINGS{9180550,
author={Banerjee, Utsav and Das, Siddharth and Chandrakasan, Anantha P.},
booktitle={2020 IEEE International Symposium on Circuits and Systems (ISCAS)},
title={{Accelerating Post-Quantum Cryptography using an Energy-Efficient TLS Crypto-Processor}},
year={2020},
volume={},
number={},
pages={1-5},
doi={10.1109/ISCAS45731.2020.9180550}}

@misc{keccakVHDL,
  author =       "Guido Bertoni and Joan Daemen and Micheal Peeters and Gilles Van Assche and Van Keer Ronny",
  title =        {{FIPS PUB 202 - SHA-3 Standard: Permutation-Based Hash and Extendable-Output Functions}},
  year = 2015,
  url =          "https://keccak.team/hardware.html",
}

@misc{cryptoeprint:2020/049,
author = {Erdem Alkim and Hülya Evkan and Norman Lahr and Ruben Niederhagen and Richard Petri},
title = {{ISA Extensions for Finite Field Arithmetic - Accelerating Kyber and NewHope on RISC-V}},
howpublished = {Cryptology {ePrint} Archive, Paper 2020/049},
year = {2020},
url = {https://eprint.iacr.org/2020/049}
}

@techreport{SCA22,
  title        = {{RISC‑V Cryptographic Extension Proposals, Technical Report Volume I: Scalar \& Entropy Source Instructions}},
  author       = {{RISC‑V Crypto}},
  institution  = {GitHub},
  type         = {Technical Report},
  number       = {Version 1.0.1},
  year         = {2022},
  url          = {https://github.com/riscv/riscv-crypto}
}

@ARTICLE{9773945,
author={Kuo, Yao-Ming and García-Herrero, Francisco and Ruano, Oscar and Maestro, Juan Antonio},
journal={IEEE Transactions on Computers},
title={{RISC-V Galois Field ISA Extension for Non-Binary Error-Correction Codes and Classical and Post-Quantum Cryptography}},
year={2023},
volume={72},
number={3},
pages={682-692},
doi={10.1109/TC.2022.3174587}}

@misc{cryptoeprint:2025/601,
      author = {Antonio Ras and Antoine Loiseau and Mikaël Carmona and Simon Pontié and Guénaël Renault and Benjamin Smith and Emanuele Valea},
      title = {{PHOENIX: Crypto-Agile Hardware Sharing for ML-KEM and HQC}},
      howpublished = {Cryptology {ePrint} Archive, Paper 2025/601},
      year = {2025},
      url = {https://eprint.iacr.org/2025/601}
}

@ARTICLE{10294284,
author={Li, Lu and Qin, Guofeng and Yu, Yang and Wang, Weijia},
journal={IEEE Transactions on Computer-Aided Design of Integrated Circuits and Systems},
title={{Compact Instruction Set Extensions for Kyber}},
year={2024},
volume={43},
number={3},
pages={756-760},
doi={10.1109/TCAD.2023.3327104}}

@INPROCEEDINGS{10830991,
author={Zhang, Shengnan and Zhao, Yifan and Yu, Xinglong and Han, Jun},
booktitle={2024 IEEE 17th International Conference on Solid-State \& Integrated Circuit Technology (ICSICT)},
title={{RISC-V Domain-Specific Processor for Accelerating SPHINCS+ on Multi-Core Architecture}},
year={2024},
volume={},
number={},
pages={1-3},
doi={10.1109/ICSICT62049.2024.10830991}}

@Article{s23239408,AUTHOR = {Lee, Jihye and Kim, Whijin and Kim, Ji-Hoon},TITLE = {{A Programmable Crypto-Processor for National Institute of Standards and Technology Post-Quantum Cryptography Standardization Based on the RISC-V Architecture}},JOURNAL = {Sensors},VOLUME = {23},YEAR = {2023},NUMBER = {23},ARTICLE-NUMBER = {9408},URL = {https://www.mdpi.com/1424-8220/23/23/9408},PubMedID = {38067782},ISSN = {1424-8220},DOI = {10.3390/s23239408}}

@inproceedings{PQC_OpenTitan,
author = {Stelzer, Tobias and Oberhansl, Felix and Schupp, Jonas and Karl, Patrick},
title = {{Enabling Lattice-Based Post-Quantum Cryptography on the OpenTitan Platform}},
year = {2023},
isbn = {9798400702624},
publisher = {Association for Computing Machinery},
address = {New York, NY, USA},
url = {https://doi.org/10.1145/3605769.3623993},
doi = {10.1145/3605769.3623993},
booktitle = {Proceedings of the 2023 Workshop on Attacks and Solutions in Hardware Security},
pages = {51–60},
numpages = {10},
location = {Copenhagen, Denmark},
series = {ASHES '23}
}

@ARTICLE{10723802,
author={Ji, Xinyi and Dong, Jiankuo and Huang, Junhao and Yuan, Zhijian and Dai, Wangchen and Xiao, Fu and Lin, Jingqiang},
journal={IEEE Transactions on Computers},
title={{ECO-CRYSTALS: Efficient Cryptography CRYSTALS on Standard RISC-V ISA}},
year={2025},
volume={74},
number={2},
pages={401-413},
doi={10.1109/TC.2024.3483631}}

@article{Sphincs_hashes,
author = {Lopez-Valdivieso, Jonathan and Cumplido, Rene},
title = {{Design and Implementation of Hardware-Software Architecture Based on Hashes for SPHINCS+}},
year = {2024},
issue_date = {December 2024},
publisher = {Association for Computing Machinery},
address = {New York, NY, USA},
volume = {17},
number = {4},
issn = {1936-7406},
url = {https://doi.org/10.1145/3653459},
doi = {10.1145/3653459},
journal = {ACM Trans. Reconfigurable Technol. Syst.},
month = oct,
articleno = {54},
numpages = {22}
}

@misc{cryptoeprint:2024/1515,
      author = {Jipeng Zhang and Yuxing Yan and Junhao Huang and Cetin Kaya Koc},
      title = {{Optimized Software Implementation of Keccak, Kyber, and Dilithium on RV32,64IMBV}},
      howpublished = {Cryptology {ePrint} Archive, Paper 2024/1515},
      year = {2024},
      url = {https://eprint.iacr.org/2024/1515}
}

@misc{cryptoeprint:2024/1192,
      author = {Amin Abdulrahman and Felix Oberhansl and Hoang Nguyen Hien Pham and Jade Philipoom and Peter Schwabe and Tobias Stelzer and Andreas Zankl},
      title = {{Towards ML-KEM \& ML-DSA on OpenTitan}},
      howpublished = {Cryptology {ePrint} Archive, Paper 2024/1192},
      year = {2024},
      doi = {10.1109/SP61157.2025.00220},
      url = {https://eprint.iacr.org/2024/1192}
}

@ARTICLE{9946370,
author={Aikata, Aikata and Mert, Ahmet Can and Imran, Malik and Pagliarini, Samuel and Roy, Sujoy Sinha},
journal={IEEE Transactions on Circuits and Systems I: Regular Papers},
title={{KaLi: A Crystal for Post-Quantum Security Using Kyber and Dilithium}},
year={2023},
volume={70},
number={2},
pages={747-758},
doi={10.1109/TCSI.2022.3219555}}

@article{Wang_Zhang_Zhang_Gu_Cao_2024, title={{Optimized Hardware-Software Co-Design for Kyber and Dilithium on RISC-V SoC FPGA}}, volume={2024},  url={https://tches.iacr.org/index.php/TCHES/article/view/11671},  DOI={10.46586/tches.v2024.i3.99-135}, number={3}, journal={IACR Transactions on Cryptographic Hardware and Embedded Systems}, author={Wang, Tengfei and Zhang, Chi marriage Zhang, Xiaolin and Gu, Dawu and Cao, Pei}, year={2024}, month={Jul.}, pages={99–135} }

@INPROCEEDINGS{10993202,
  author={Dolmeta, Alessandra and Di Matteo, Stefano and Valea, Emanuele and Carmona, Mikael and Loiseau, Antoine and Martina, Maurizio and Masera, Guido},
  booktitle={2025 Design, Automation \& Test in Europe Conference (DATE)}, 
  title={{TYRCA: A RISC-V Tightly-Coupled Accelerator for Code-Based Cryptography}}, 
  year={2025},
  volume={},
  number={},
  pages={1-7},
  doi={10.23919/DATE64628.2025.10993202}}

@techreport{nist_cswp_39,
  author      = {Elaine Barker and Lily Chen and David Cooper and Dustin Moody and Andrew Regenscheid and Murugiah Souppaya and Bill Newhouse and Russ Housley and Sean Turner and William Barker and Karen Scarfone},
  title       = {{Considerations for Achieving Crypto Agility: Strategies and Practices}},
  institution = {National Institute of Standards and Technology},
  year        = {2025},
  type        = {NIST Cybersecurity White Paper (CSWP)},
  number      = {NIST CSWP 39 2pd},
  address     = {Gaithersburg, MD},
  doi         = {10.6028/NIST.CSWP.39.2pd},
  url         = {https://doi.org/10.6028/NIST.CSWP.39.2pd}
}

@article{SAKWA2026100975,
title = {{Survey on post-quantum cryptography implementations and deployment challenges}},
journal = {Computer Science Review},
volume = {61},
pages = {100975},
year = {2026},
issn = {1574-0137},
doi = {https://doi.org/10.1016/j.cosrev.2026.100975},
url = {https://www.sciencedirect.com/science/article/pii/S1574013726000833},
author = {Cyprian Omukhwaya Sakwa and Fagen Li},
}

@INPROCEEDINGS{11043433,
  author={Dolmeta, Alessandra and Piscopo, Valeria and Mirigaldi, Mattia and Martina, Maurizio and Masera, Guido},
  booktitle={2025 IEEE International Symposium on Circuits and Systems (ISCAS)}, 
  title={{RISC-V Based Keccak Co-Processor for NIST Post-Quantum Cryptography Standards}}, 
  year={2025},
  volume={},
  number={},
  pages={1-5},
  doi={10.1109/ISCAS56072.2025.11043433}}

@inproceedings{10.1145/3706594.3726982,
author = {Waucquez, Luis and Rodriguez, Alfonso},
title = {{EROS: Extensible Reliable Offloading Solution}},
year = {2025},
isbn = {9798400713934},
publisher = {Association for Computing Machinery},
address = {New York, NY, USA},
url = {https://doi.org/10.1145/3706594.3726982},
doi = {10.1145/3706594.3726982},
booktitle = {Proceedings of the 22nd ACM International Conference on Computing Frontiers: Workshops and Special Sessions},
pages = {82–85},
numpages = {4},
location = {
},
series = {CF '25 Companion}
}

@INPROCEEDINGS{10752475,
  author={Hepola, Kari and Arachchige, Tharaka Ranasinghe and Multanen, Joonas and Jääskeläinen, Pekka},
  booktitle={2024 IEEE Nordic Circuits and Systems Conference (NorCAS)}, 
  title={{Fully Automatic Compiler Retargeting and CV-X-IF Hardware Interface Generation for RISC-V Custom Instructions}}, 
  year={2024},
  volume={},
  number={},
  pages={1-7},
  doi={10.1109/NorCAS64408.2024.10752475}}

@Article{electronics14153152,
AUTHOR = {Popovici, Cosmin-Andrei and Stan, Andrei and Botezatu, Nicolae-Alexandru and Manta, Vasile-Ion},
TITLE = {{RiscADA: RISC-V Extension for Optimized Control of External D/A and A/D Converters}},
JOURNAL = {Electronics},
VOLUME = {14},
YEAR = {2025},
NUMBER = {15},
ARTICLE-NUMBER = {3152},
URL = {https://www.mdpi.com/2079-9292/14/15/3152},
ISSN = {2079-9292},
DOI = {10.3390/electronics14153152}
}

@ARTICLE{11414189,
  author={Seo, Seog Chung and Kim, YoungBeom},
  journal={IEEE Transactions on Circuits and Systems II: Express Briefs}, 
  title={{A Hardware/Software Co-Optimization of HQC Using Tightly-Coupled Accelerators on a 32-bit Ibex Core}}, 
  year={2026},
  volume={},
  number={},
  pages={1-1},
  doi={10.1109/TCSII.2026.3668482}}

@misc{cryptoeprint:2023/1505,
author = {Konstantina Miteloudi and Joppe Bos and Olivier Bronchain and Björn Fay and Joost Renes},
title = {{PQ.v.ALU.e: Post-quantum RISC-v custom ALU extensions on dilithium and kyber}},
howpublished = {Cryptology {ePrint} Archive, Paper 2023/1505},
year = {2023},
url = {https://eprint.iacr.org/2023/1505}
}

@misc{cryptoeprint:2024/367,
      author = {Markku-Juhani O. Saarinen},
      title = {{Accelerating SLH-DSA by Two Orders of Magnitude with a Single Hash Unit}},
      howpublished = {Cryptology {ePrint} Archive, Paper 2024/367},
      year = {2024},
      doi = {10.1007/978-3-031-68376-3_9},
      url = {https://eprint.iacr.org/2024/367}
}

@article{10.1145/3643826,
author = {Li, Lu and Tian, Qi and Qin, Guofeng and Chen, Shuaiyu and Wang, Weijia},
title = {{Compact Instruction Set Extensions for Dilithium}},
year = {2024},
issue_date = {March 2024},
publisher = {Association for Computing Machinery},
address = {New York, NY, USA},
volume = {23},
number = {2},
issn = {1539-9087},
url = {https://doi.org/10.1145/3643826},
doi = {10.1145/3643826},
journal = {ACM Trans. Embed. Comput. Syst.},
month = mar,
articleno = {23},
numpages = {21}
}

@INPROCEEDINGS{9217834,
  author={Amiet, Dorian and Leuenberger, Lukas and Curiger, Andreas and Zbinden, Paul},
  booktitle={2020 23rd Euromicro Conference on Digital System Design (DSD)}, 
  title={{FPGA-based SPHINCS+ Implementations: Mind the Glitch}}, 
  year={2020},
  volume={},
  number={},
  pages={229-237},
  doi={10.1109/DSD51259.2020.00046}}

@INPROCEEDINGS{10817843,
  author={Awakiahra, Masakazu and Yoshimoto, Keiji and Kanzawa, Shota and Miyoshi, Akane and Nogami, Yasuyuki and Kodera, Yuta},
  booktitle={2024 Twelfth International Symposium on Computing and Networking Workshops (CANDARW)}, 
  title={{Proposal and Evaluation for Efficient splitfft and mergefft in FALCON}}, 
  year={2024},
  volume={},
  number={},
  pages={247-251},
  doi={10.1109/CANDARW64572.2024.00047}}

@INPROCEEDINGS{9460703,
  author={Berthet, Quentin and Upegui, Andres and Gantel, Laurent and Duc, Alexandre and Traverso, Giulia},
  booktitle={2021 IEEE International Parallel and Distributed Processing Symposium Workshops (IPDPSW)}, 
  title={{An Area-Efficient SPHINCS+ Post-Quantum Signature Coprocessor}}, 
  year={2021},
  volume={},
  number={},
  pages={180-187},
  doi={10.1109/IPDPSW52791.2021.00034}}

@ARTICLE{10776955,
  author={Dolmeta, Alessandra and Martina, Maurizio and Masera, Guido},
  journal={IEEE Access}, 
  title={{ATHOS: A Hybrid Accelerator for PQC CRYSTALS-Algorithms Exploiting New CV-X-IF Interface}}, 
  year={2024},
  volume={12},
  number={},
  pages={182340-182352},
  doi={10.1109/ACCESS.2024.3511340}}

@article{10.1145/3579092,
author = {Karl, Patrick and Schupp, Jonas and Fritzmann, Tim and Sigl, Georg},
title = {Post-Quantum Signatures on RISC-V with Hardware Acceleration},
year = {2024},
issue_date = {March 2024},
publisher = {Association for Computing Machinery},
address = {New York, NY, USA},
volume = {23},
number = {2},
issn = {1539-9087},
url = {https://doi.org/10.1145/3579092},
doi = {10.1145/3579092},
journal = {ACM Trans. Embed. Comput. Syst.},
month = mar,
articleno = {30},
numpages = {23}
}

@INPROCEEDINGS{9609917,
  author={Beckwith, Luke and Nguyen, Duc Tri and Gaj, Kris},
  booktitle={2021 International Conference on Field-Programmable Technology (ICFPT)}, 
  title={{High-Performance Hardware Implementation of CRYSTALS-Dilithium}}, 
  year={2021},
  volume={},
  number={},
  pages={1-10},
  doi={10.1109/ICFPT52863.2021.9609917}}

@misc{cryptoeprint:2023/1885,
      author = {Michael Schmid and Dorian Amiet and Jan Wendler and Paul Zbinden and Tao Wei},
      title = {{Falcon Takes Off - A Hardware Implementation of the Falcon Signature Scheme}},
      howpublished = {Cryptology {ePrint} Archive, Paper 2023/1885},
      year = {2023},
      url = {https://eprint.iacr.org/2023/1885}
}

@misc{cryptoeprint:2019/844,
      author = {Matthias J.  Kannwischer and Joost Rijneveld and Peter Schwabe and Ko Stoffelen},
      title = {{pqm4: Testing and Benchmarking {NIST} {PQC} on {ARM} Cortex-M4}},
      howpublished = {Cryptology {ePrint} Archive, Paper 2019/844},
      year = {2019},
      url = {https://eprint.iacr.org/2019/844}
}

@InProceedings{10.1007/978-981-95-8399-7_12,
author="Hu, Zhiming
and Liu, Tianlin
and Wang, Xinhua
and Li, Sizhao
and Fu, Xiaojing
and Li, Qiuliang",
editor="Liu, Huazhong
and Ibrahim, Shadi
and Rauber, Thomas",
title="KiRa: A Unified Memory Architecture for Efficient Post-quantum Cryptographic Algorithms",
booktitle="Algorithms and Architectures for Parallel Processing",
year="2026",
publisher="Springer Nature Singapore",
address="Singapore",
pages="210--229",
isbn="978-981-95-8399-7"
}

\begin{IEEEbiography}[{\includegraphics[width=1in,height=1.25in,clip,keepaspectratio]{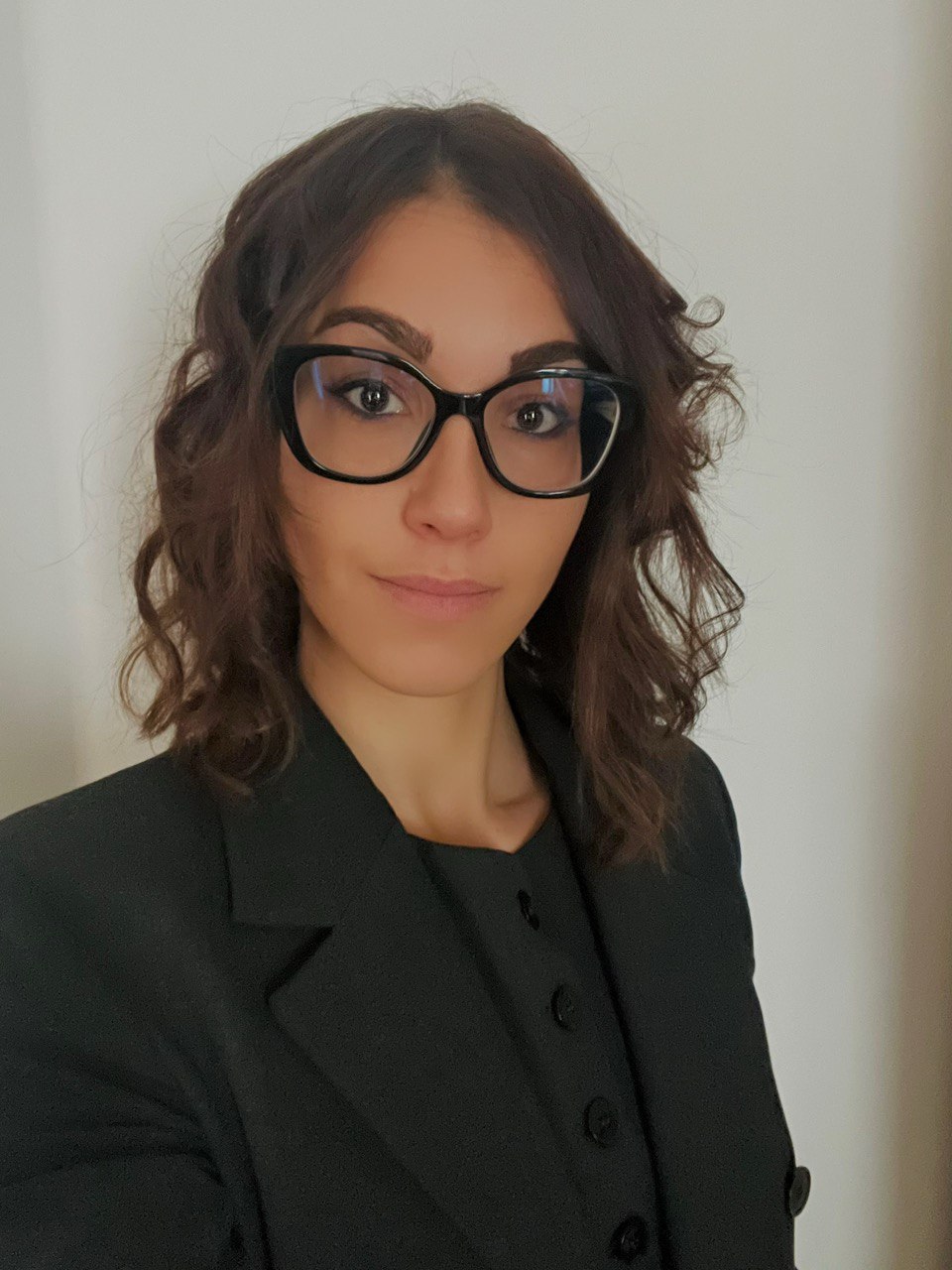}}]{Alessandra Dolmeta}
received the M.Sc. degree in Electronic Engineering and the Ph.D. degree in Electronics and Communications Engineering from Politecnico di Torino, Italy, in 2022 and 2025, respectively. Since 2026, she has been a Postdoctoral Researcher with the Department of Electronics and Telecommunications, Politecnico di Torino. Her research interests include computer architecture, hardware security, post-quantum cryptography, and the design of specialized hardware accelerators for secure and low-power embedded systems. Her current research focuses on hardware-software co-design for the RISC-V ecosystem, custom instruction set extensions, and quantum-resistant cryptographic accelerators.
\end{IEEEbiography}

\begin{IEEEbiography}[{\includegraphics[width=1in,height=1.25in,clip,keepaspectratio]{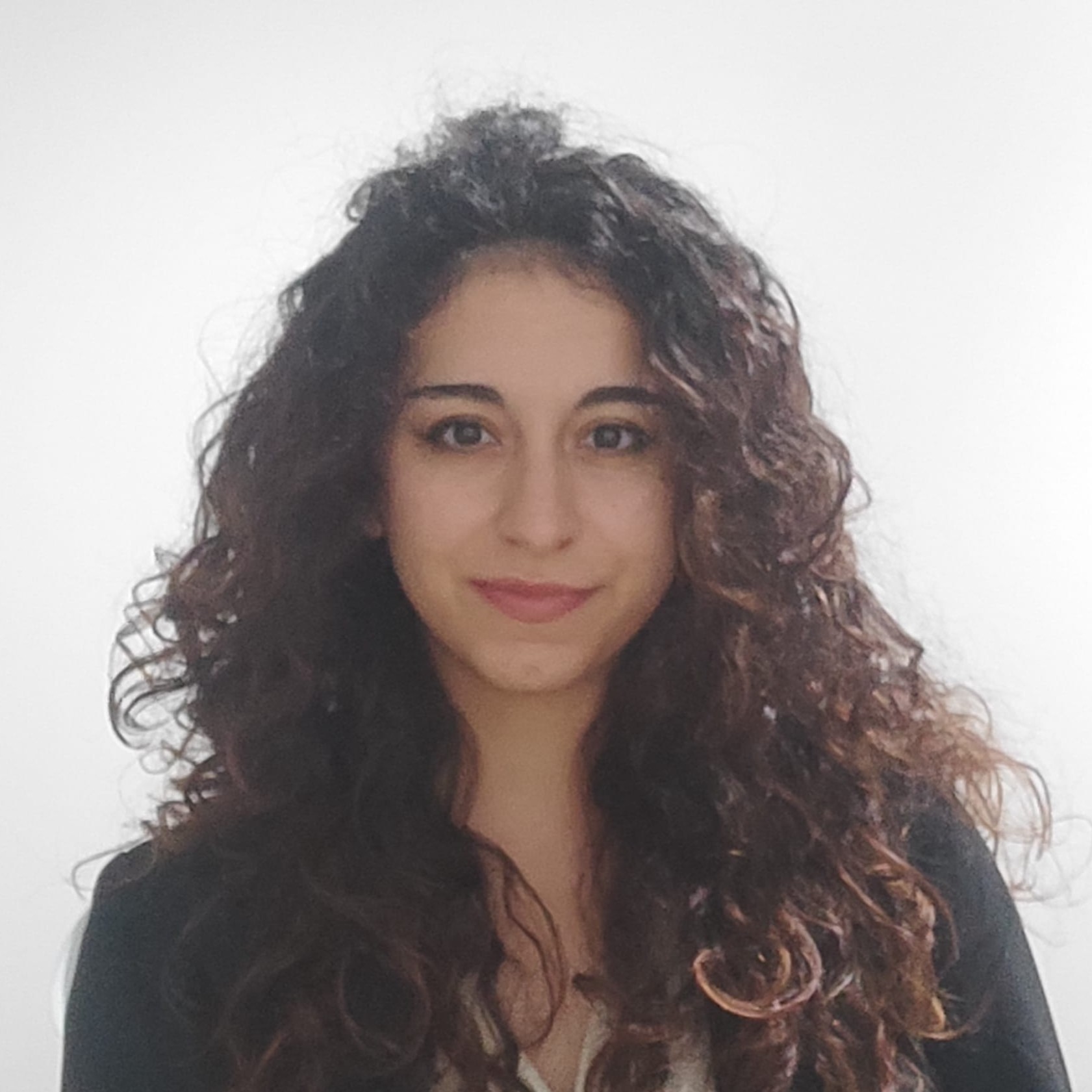}}]{Valeria Piscopo} received the M.Sc. degree in Electronic Engineering from Politecnico di Torino, Italy, in 2024. She is currently a second-year Ph.D. candidate in Electronics and Communications Engineering at Politecnico di Torino. Her research focuses on the design and optimization of specialized cryptographic hardware accelerators for secure and energy-efficient embedded systems, with particular emphasis on the RISC-V ecosystem.

\end{IEEEbiography}

\begin{IEEEbiography}[{\includegraphics[width=1in,height=1.25in,clip,keepaspectratio]{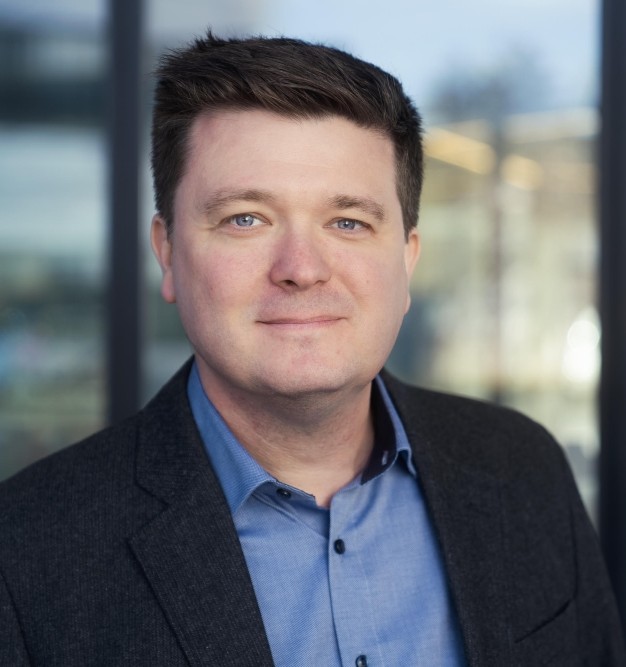}}]{Michael Hutter}
s Full Professor at the University of the Bundeswehr Munich, Germany, where he leads the Embedded Systems Security (ESSEC) research group within the Department of Computer Science and the Research Institute CODE. He received his PhD and Habilitation (venia docendi) in Applied Cryptography from the Institute of Information Security (ISEC, formerly IAIK) at Graz University of Technology, Austria. Prior to academia, he spent more than a decade in industry, serving as Head of Innovation and Concepts at PQShield (Oxford, UK) and as Senior Principal Engineer at Rambus Inc. (former CRI, USA). He holds over 30 international patents and has co-authored more than 70 peer-reviewed publications. Prof. Hutter has served on numerous program committees of leading conferences and as Editor-in-Chief of TCHES 2025. He is currently a member of the steering committee of CHES and Austrochip, and recently contributed as DT5 Topic Chair at DATE 2024–2026.
\end{IEEEbiography}

\begin{IEEEbiography}[{\includegraphics[width=1in,height=1.25in,clip,keepaspectratio]{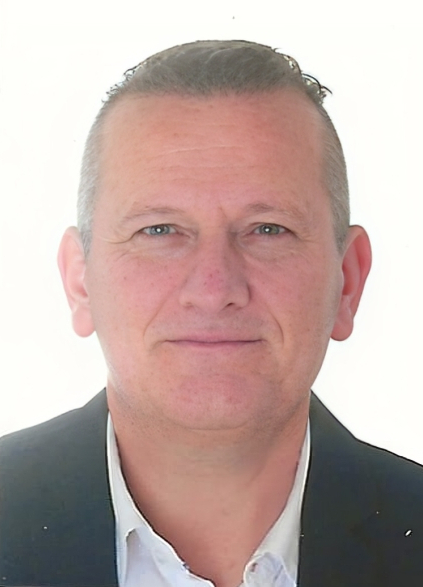}}]{Maurizio Martina}
Maurizio Martina received the M.Sc. and Ph.D. degrees in Electronic Engineering and Electronic and Communications Engineering from Politecnico di Torino, Italy, in 2000 and 2004, respectively. He is currently a Professor with the Department of Electronics 
and Telecommunications, Politecnico di Torino.
His research interests include computer architecture and VLSI design of digital integrated circuits for image and video coding, forward error correction, cryptography, and artificial intelligence. He has authored or co‑authored more than 200 scientific 
publications and holds two patents. He served as Associate Editor of the IEEE Transactions on Circuits and Systems—I (2018–2022) and as Guest Editor of several special issues, including the BioCAS 2017 special issue in IEEE Transactions on Biomedical Circuits and Systems and the ISCAS 2023 special issue in IEEE Transactions on Circuits and Systems—II. He has been a member of the organizing and technical committees of several IEEE conferences, including BioCAS 2017, ICECS 2019, AICAS 2020, and PRIME since 2020.
\end{IEEEbiography}

\begin{IEEEbiography}[{\includegraphics[width=1in,height=1.25in,clip,keepaspectratio]{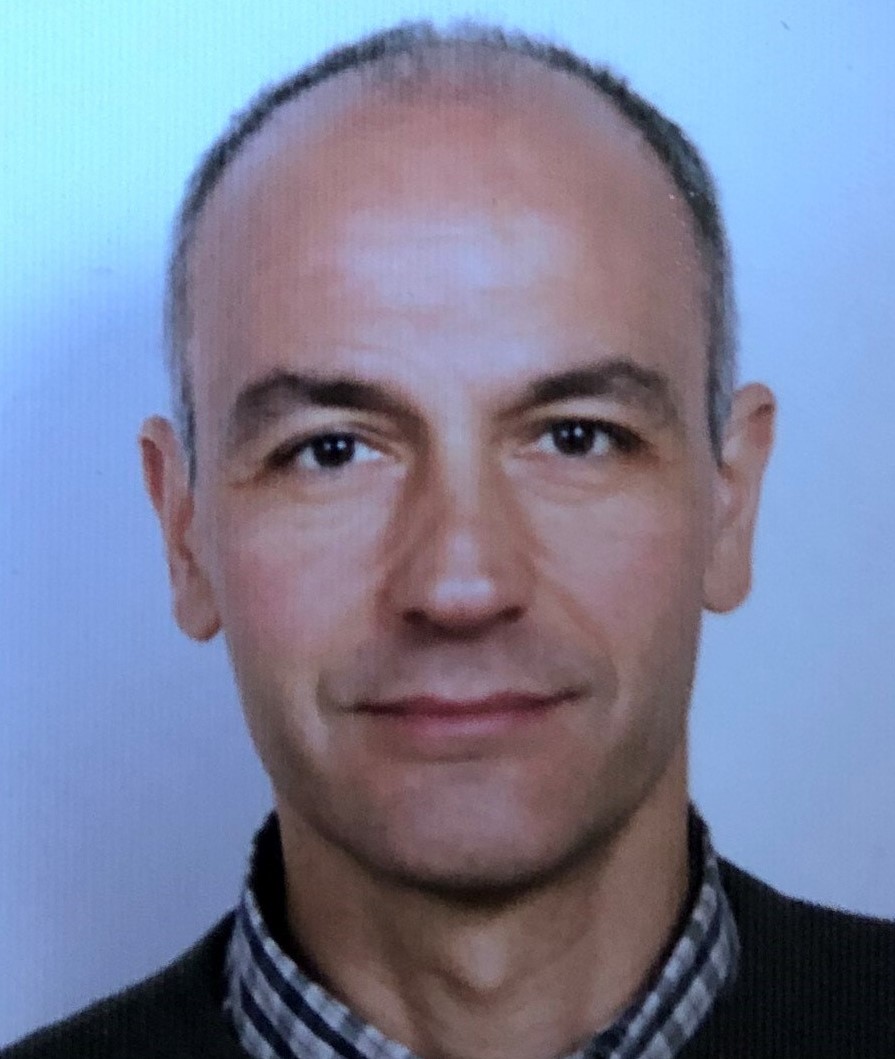}}]{Guido Masera}
 (SM’07) received the Dr.-Ing. (summa cum laude) and Ph.D. degrees in electronic engineering from Politecnico di Torino, Italy. He is a Professor with the Electronics and Telecommunications Department, Politecnico di Torino, since 1992. His research interests include several aspects in the design of digital integrated circuits and systems, with a special emphasis on high-performance architectures for communications, forward error correction, image and video coding, cryptography and hardware accelerators for machine learning. He has more than 200 publications, two patents and was a designer of several ASIC components. Dr. Masera is an Associate Editor of MDPI Electronics and a former Associate Editor of the IEEE Transactions on Circuits and Systems I, IEEE Transactions on Circuits and Systems II and the IET Circuits, Devices \& Systems. 
\end{IEEEbiography}

\end{document}